\documentclass[aps,amsmath,amssymb,floatfix,superscriptaddress,showkeys]{revtex4}

\pdfoutput=1
\usepackage{color}
\usepackage{graphicx}
 
\renewcommand{\thesection}{\arabic{section}}
\renewcommand{\thesubsection}{\arabic{section}.\arabic{subsection}}

\begin{document}

\title{Bayesian gradient sensing in the presence of rotational diffusion} 

\author{Maja Novak}
\affiliation{TU Dresden, Dresden, Germany}
\author{Benjamin M. Friedrich}
\affiliation{TU Dresden, Dresden, Germany}
\email{benjamin.m.friedrich@tu-dresden.de}

\date{\today}


\begin{abstract} 
Biological cells estimate concentration gradients of signaling molecules with a precision that 
is limited not only by sensing noise, but additionally by the cell's own stochastic motion.
We ask for the theoretical limits of gradient estimation in the presence of both motility and sensing noise.
We introduce a minimal model of a stationary chemotactic agent in the plane subject to rotational diffusion, 
which uses Bayesian estimation to optimally infer a gradient direction from noisy concentration measurements.
Contrary to the known case of gradient sensing by temporal comparison, 
we show that for spatial comparison, 
the ultimate precision of gradient sensing scales not with the rotational diffusion time, 
but with its square-root.
To achieve this precision, 
an individual agent needs to know its own rotational diffusion coefficient.
This agent can accurately estimate the expected variability within an ensemble of agents. 
If an agent, however, does not account for its own motility noise, 
Bayesian estimation fails in a characteristic manner.
\end{abstract}

\newcommand{\g}{\mathbf{g}} 
\newcommand{\h}{\mathbf{h}} 
\renewcommand{\L}{\mathcal{L}} 
\newcommand{\M}{\mathcal{M}} 
\newcommand{\N}{\mathcal{N}} 
\renewcommand{\S}{\mathcal{S}} 
\newcommand{\R}{\mathbf{R}} 
\renewcommand{\r}{\mathbf{r}} 
\newcommand{\T}{\mathcal{T}} 
\renewcommand{\O}{\mathcal{O}} 
\renewcommand{\v}{\mathbf{v}} 
\newcommand{\micron}{\ensuremath{\mu}\mathrm{m}} 
\newcommand{\vv}{\mathbf{v}} 
\newcommand{\CV}{\mathrm{CV}} 
\newcommand{\SNR}{\mathrm{SNR}} 
\newcommand{\VM}{\mathcal{VM}} 
\newcommand{\WN}{\mathcal{WN}} 

\newcommand{\wt}[1]{\widetilde{#1}}
\newcommand{\wh}[1]{\widehat{#1}}
\renewcommand{\ol}[1]{\overline{#1}}
\newcommand{\true}[1]{ {#1_\mathrm{true}} }

\maketitle

\section{Introduction}

Many motile biological cells navigate in concentration gradients of signaling molecules in a process termed chemotaxis
\cite{Berg1972,Eisenbach2006,Alvarez2014,Devreotes1988,Zigmond1977,Gregor2007}.
At cellular scales, 
the stochastic binding of signaling molecules results in molecular shot noise and
renders concentration measurements inherently noisy \cite{Berg1977}.
This sensing noise imposes physical limits on the precision of chemotaxis
\cite{Berg1977,Bialek2005,Kaizu2014,Rappel2008,Hu2010,Endres2008,tenWolde2016}.
Experimental work suggests that biological cells indeed operate at these limits in shallow concentration gradients
\cite{vanHaastert2007,Mortimer2009,Fuller2010,Amselem2012,Brumley2019}. 
Temporal averaging over extended measurement intervals
is a common strategy to reduce sensing noise \cite{Berg1977,Segall1986,Kashikar2012,Hathcock2016}.
Yet, temporal averaging may be limited in time-varying environments 
\cite{Celani2010,Aquino2014,Hein2016,Brumley2019},
or more directly by the stochastic motion of chemotactic agents themselves \cite{Strong1998}.
How should a chemotactic agent integrate previous and more recent measurements
to most accurately estimate the relative direction of a concentration gradient
if this direction changes stochastically in time?

Previous work suggested that bacteria use Kalman filters to track a time-dependent concentration signal, 
providing an optimal weighting of past and recent measurements \cite{Andrews2006},
see also \cite{Strong1998}.
Yet, Kalman filters address linear problems \cite{Kalman1960}, while sensing a direction is a nonlinear problem of circular statistics, 
which prompts Bayesian updating of an angular likelihood distribution at each time step \cite{Melsa1978}.
Bayesian estimation had been successfully applied, e.g., for
monitoring time-varying environments with two states \cite{Kobayashi2010},
estimating absolute concentration \cite{Endres2009,Zechner2016,Scholz2017},
or temporal changes thereof \cite{Mora2010,Hein2016}, as well as
classification tasks \cite{Libby2007,Siggia2013}, and even
decision making in the immune system \cite{Mayer2019}. 

The specific problem of Bayesian sensing of direction was addressed previously
\cite{
Endres2008,
Mortimer2009,
Hu2010, 
Hu2011}, 
yet without considering motility noise.
Likewise, 
the infotaxis algorithm, which computes a likelihood map for the position of a hidden target, 
does not include motility noise \cite{Vergassola2007}.
In the engineering literature,
directional sensing is known as `bearing tracking', 
and estimation algorithms do indeed account for motility noise \cite{Markovic2012,Kurz2016}.
However, to the best of our knowledge, 
an analytical theory of optimal directional sensing that accounts for motility noise is missing.

Here, we derive theoretical limits for the precision of gradient sensing by chemotactic agents such as biological cells in the presence of both sensing and motility noise.
We consider a minimal model of a chemotactic agent in the plane that attempts to track 
the direction of an external concentration gradient.
The agent performs noisy concentration measurements, 
while it undergoes rotational diffusion.
This agent integrates subsequent measurements into a likelihood distribution of possible gradients using Bayesian updating.

Our manuscript is structured as follows:
We first briefly review Bayesian gradient sensing without motility noise to introduce notation.
We recapitulate how temporal averaging improves the precision of gradient estimates as a function of measurement time.
We then introduce motility noise and consider an agent subject to rotational diffusion.
This agent, however, is first not aware of its own motility noise,
which results in grossly erroneous gradient estimates.
In contrast, as our main result,
we show how an agent that only knows its own rotational diffusion coefficient $D$
can obtain optimal estimates of gradient direction, as well as a self-consistent estimate of the accuracy of this estimate, 
i.e., the expected dispersion in an \textit{ensemble} of agents,
which scales as $D^{-1/2}$.
This optimal gradient-sensing strategy corresponds to temporal averaging 
over a time span that likewise scales as $D^{-1/2}$.
We discuss why this new result for gradient-sensing by \textit{spatial comparison} 
is different from previous results for chemotaxis by \textit{temporal comparison}, 
which predicted a substantially longer optimal time span of temporal averaging that scales as $D^{-1}$ \cite{Strong1998}.

\section{Minimal model}

\begin{figure}
\includegraphics[width=0.8\textwidth]{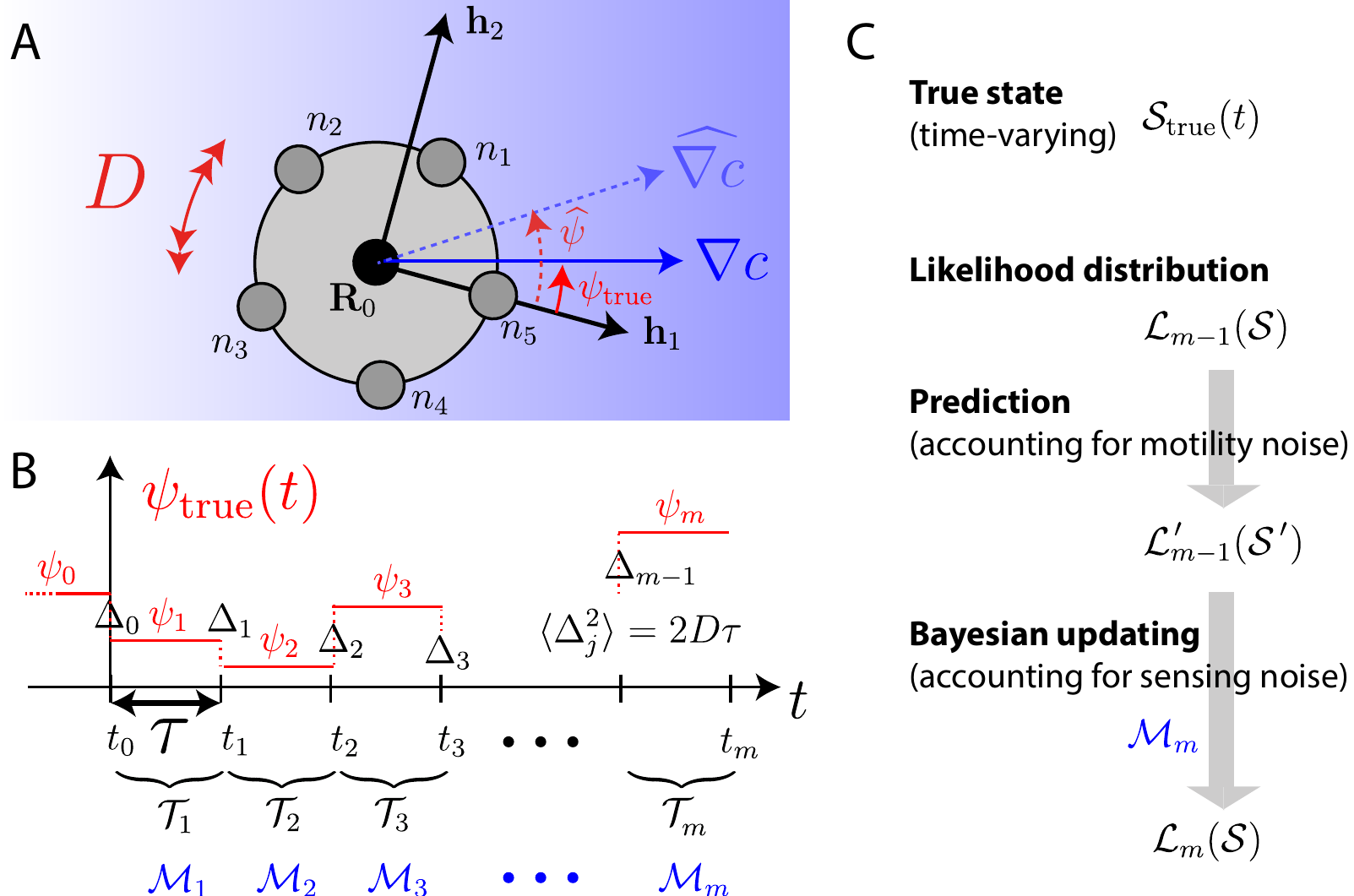}
\caption{
\textbf{Chemotactic agent subject to rotational diffusion.}
\textbf{A.}
In our minimal model, a chemotactic agents seeks to estimate an external concentration gradient $\nabla c$ of signaling molecules
(relative to its material frame $\h_1$ and $\h_2$)
by counting binding events at $N$ sensor sites spaced equidistantly on the agent's circumference.
During each measurement interval $\T_j$ of duration $\tau$, the agent obtains molecule counts $n_1,\ldots,n_{N}$,
which are combined into a (Fourier-transformed) measurement vector $\M_j$.
Between measurements, the agent is subject to rotational diffusion with rotational diffusion coefficient $D$.
\textbf{B.}
The angle $\psi_\mathrm{true}(t)$ enclosed between gradient direction $\nabla c$ and material frame vector $\h_1$ 
becomes a random walk with stochastic increments $\Delta_j$.
This motility noise limits the precision 
of the gradient estimate $\wh{\nabla c}$ and its direction angle $\wh{\psi}$ as estimated by the agent. 
\textbf{C.}
The relative direction of the concentration gradient represents a time-dependent state of the environment,
$\S_\mathrm{true}(t) = S_j$ for $t\in\T_j$ where $S_j=(c_0,\alpha_0,\psi_j)$.
The agent computes a likelihood distribution $\L(\S)$ of possible concentration gradients, iteratively executing
a prediction step that accounts for its rotational diffusion (which flattens the distribution), and 
an update step that incorporates a new measurement $\M_m$ (which usually sharpens the distribution).
}
\label{figure1}
\end{figure}

We consider a chemotactic agent in the plane, see Fig.~\ref{figure1}.
Orthonormal vectors $\h_1$ and $\h_2$ define its material frame.
The agent is subject to rotational diffusion
with effective diffusion coefficient $D=D_\mathrm{rot}$ (with units of inverse time), i.e., 
$\langle \h_1(t_0)\cdot\h_1(t_0+t)\rangle=\exp(-D|t|)$.
For simplicity, the agent is stationary with center position $\R_0$.

The agent seeks to estimate the strength and direction 
of an external concentration gradient
\begin{equation}
\label{eq:c}
c(\r)=c_0\,
\left[
1+\frac{\alpha_0}{a}\,\g\cdot(\r-\R_0)
\right]
\end{equation}
with 
\textit{base concentration} $c_0$, 
dimensionless \textit{gradient strength} $\alpha_0=|\nabla c| a / c_0$
(normalized by a sensing length-scale $a$ set by the dimensions of the agent), and
gradient direction vector
$\g = \cos\psi_0\,\h_1+\sin\psi_0\,\h_2$ of unit length,
such that $\psi_0$ denotes the \textit{gradient angle} enclosed by $\h_1$ and $\g$.
The concentration gradient Eq.~(\ref{eq:c}) represents an unknown \textit{state of the environment}, 
$\true{\S}(t) = \left(c_0,\alpha_0,\true{\psi}(t) \right)$, 
where $\true{\psi}(0) = \psi_0$ at time $t_0=0$.
The agent shall be equipped with $N$ sensors 
placed equidistantly on the circumference of a disk of radius $a$
at respective positions 
$\r_j = \R_0 + a\, \cos(2\pi j/N)\,\h_1 + a\, \sin(2\pi j/N)\,\h_2$, 
see Fig.~\ref{figure1}A.
To simplify future calculations, 
we introduce complex vector notation $\h=\h_1+i\h_2$,
which gives,
$\g=\mathrm{Re}\,e^{i\psi_0}\h^\ast$ and
$\r_j = \R_0 + \mathrm{Re}\, \eta^j \h^\ast$, 
where
$\eta = \exp(2\pi i /N)$ denotes the $N^\mathrm{th}$ root of unity.
Each sensor detects stochastic binding events of molecules with rate
$\Lambda_j = \lambda c_j / N$
proportional to local concentration
$c_j = c(\r_j) = c_0 [1+\alpha_0 \cos(2\pi j/N-\psi_0)]$
for $j=1,\ldots,N$.
The sensor count $n_j$ during a time interval $\tau$ 
becomes a Poissonian random variable with expectation value 
$\ol{n}_j = \langle n_j \rangle = \Lambda_j \tau$ and variance $\ol{n}_j$.
As fusion of this sensor information, 
we consider its Fourier transform
$\wt{n}_k = \sum_{j=1}^{N} n_j \eta^{jk}$ for $k=0,\ldots,N-1$
(where we use the complex conjugate of the discrete Fourier transform to simplify formulas below).
In a linear concentration field as given by Eq.~(\ref{eq:c}), 
only the first two Fourier coefficients have non-zero expectation values,
$\langle \wt{n}_0 \rangle = \nu_0$, 
$\langle \wt{n}_1 \rangle = \alpha_0\nu_0 e^{i\psi_0} /2$ (for $N\ge 3$),
whereas
$\langle \wt{n}_k \rangle = 0$ for $2\le k\le N-2$.
Correspondingly, we assume that the agent stores only a measurement vector
$\M = 
\left(
\wt{n}_0, \wt{n}'_1, \wt{n}_1'' 
\right) \in \mathbb{R}^3$, 
where $\wt{n}_1=\wt{n}_1'+i\,\wt{n}_1''$ denotes decomposition into real and imaginary part.

In the following, 
we consider discrete time dynamics
with subsequent measurement intervals
$\T_j=(t_{j-1},t_j)$ of duration $\tau$
and discrete rotational diffusion events at times $t_j=j\tau$, 
where the agent rotates by a random angle $\Delta_j$ 
that is normally distributed with zero mean and variance 
$\langle \Delta_i \Delta_j \rangle = 2D\tau\,\delta_{ij}$, 
see Fig.~\ref{figure1}B.

The agents estimates the state of the environment as
$\wh{\S}=\left(\wh{c},\wh{\alpha},\wh{\psi}\right)$
based on the sequence of measurements
$\M_1,\ldots,\M_m$
taken during the time intervals $\T_1,\ldots,\T_m$
using a maximum-likelihood estimate as detailed below.

\section{Bayesian gradient sensing without motility noise} 

\subsection{The measurement process for a single measurement}

From the fact that the molecule counts $n_j$ are independent Poisson random variables,
we readily find the expectation value
$ \mu_0 = \langle \M \rangle $ 
and covariance matrix
$ \Sigma_0 = \langle (\M-\mu_0)(\M-\mu_0)^T \rangle $
of a single measurement $\M$
if the true state is $\S_0=(c_0,\alpha_0,\psi_0)$.
Interestingly, 
if the agent possesses at least four sensors, $N\ge 4$, 
both $\mu_0$ and $\Sigma_0$ are independent of the number $N$ of sensors.
For $N=2$, gradient-sensing obviously becomes impossible if $\psi_0=\pm \pi/2$, 
while the precision of gradient-sensing (weakly) depends on $\psi_0$ for $N=3$, 
i.e.\ it depends on the orientation of the agent relative to the gradient direction, see appendix \ref{app:no_impatience}.
For $N\ge 4$, we find
\begin{equation}
\label{eq:mean}
\mu_0 
= \nu_0\,
\left(
1,
\alpha_0 \cos(\psi_0)/2, 
\alpha_0 \sin(\psi_0)/2 
\right)^T 
\end{equation}
and covariance matrix
\begin{equation}
\label{eq:covar}
\Sigma_0 = 
\nu_0
\begin{pmatrix}
1 & 
\alpha_0 \cos(\psi_0)/2 &
\alpha_0 \sin(\psi_0)/2 \\
\alpha_0 \cos(\psi_0)/2 & 
1/2 & 0 \\
\alpha_0 \sin(\psi_0) / 2 & 
0 & 
1 / 2
\end{pmatrix} \quad, 
\end{equation}
where we introduced short-hand
$\nu_0=\lambda\tau c_0$, 
and
$\nu=\lambda\tau c$ for later use.
Interestingly, the covariance matrix $\Sigma_0$ possesses non-zero off-diagonal entries, 
i.e., 
$\wt{n}_0$ (``measuring absolute concentration'') and 
$\wt{n}_1$ (``measuring a gradient'') 
are not independent. 

In the limit of large molecule counts, 
$\ol{n}_j\gg 1$,
we can employ a diffusion approximation, 
and approximate the probability distribution of each $n_j$
by a normal distribution with mean $\ol{n}_j$ and variance $\ol{n}_j$.
The discrete Fourier transform is a linear transformation, 
hence the distribution of measurement vectors $\M$ can likewise be approximated as a multi-variate Gaussian, 
using the mean values and co-variance matrix computed above,
$P(\M\,|\,\S_0) = \N(\mu_0,\Sigma_0)$.

The agent, which does not know $\S_0$, 
anticipates that measurements $\M$ are distributed according to an analogous $P(\M\,|\,\S)$ for any hypothetical state $\S$, i.e.,
\begin{equation}
\label{eq:measurement_process}
P(\M\,|\,\S) 
= \N(\mu,\Sigma) 
= (2\pi)^{-3/2} |\Sigma |^{-1/2}\, \exp \left( -\frac{1}{2} (\M - \mu)^T\, \Sigma^{-1}\, (\M - \mu) \right) \quad .
\end{equation}
Here, $\mu$ and $\Sigma$ are defined analogously to Eqs.~(\ref{eq:mean}) and (\ref{eq:covar}) 
with substitutions $c_0\rightarrow c$, $\alpha_0\rightarrow\alpha$, $\psi_0\rightarrow\psi$, 
such that $\mu_0=\mu(\S_0)$ and $\Sigma_0=\Sigma(\S_0)$.

\subsection{Signal-to-noise ratio}

We introduce the signal-to-noise ratio of gradient sensing, $\SNR$
(with prefactor matching \cite{Kromer2018}), which characterizes sensing noise
\begin{align}
\SNR = 2\frac
{|\langle\wt{n}_1\rangle|^2}
{\langle|\wt{n}_1 -\langle\wt{n}_1\rangle |^2\rangle} 
= \frac{\alpha_0^2\nu_0}{2} 
\quad.
\end{align}
Here, we used Eqs.~(\ref{eq:mean}) and (\ref{eq:covar}) in the last step, 
see also appendix A.1. 
Explicitly, $\SNR = \lambda \tau |\nabla c|^2 a^2 / (2c_0)$,
i.e., the signal-to-noise ratio scales with measurement time $\tau$.
Below, we show that the $\SNR$ sets the precision of a single measurement.

\subsection{Bayesian update rule for likelihood function}

The agent computes a likelihood
$\L_m=\L(\S\,|\,\M_{1:m})$
for each possible state $\S=(c,\alpha,\psi)$ of the environment, 
based on all previous measurements $\M_1,\ldots,\M_m$.
The corresponding maximum-likelihood state estimate at time $t_m$ reads
$\wh{\S}_m = \mathrm{argmax}_\S\, \L_m$.
We are especially interested in the 
maximum-likelihood estimate $\wh{\psi}_m$ of the gradient direction, 
and the estimated precision of this estimate (quantified below in terms of a so-called measure of concentration).

After each measurement, the agent updates the likelihood function
$\L_m(\S)$, using Bayes' rule
\begin{equation}
\label{eq:Bayes_rule}
\L_{m} = \L(\S\,|\,\M_{1:m}) = \frac{ P(\M_{m}\,|\,\S)}{P(\M_{m}|\M_{1:m-1})}\,\L_{m-1} \quad.
\end{equation}
Here, 
$P(\M\,|\,\S)$
is the probability to measure $\M$ given a specific state $\S$ (\textit{measurement process}, approximated by Eq.~(\ref{eq:measurement_process})),
$P(\M_{m}\,|\,\M_{1:m-1})=\int d\S\, P(\M_{m}\,|\,\S)\,\L_{m-1}(\S\,|\,\M_{1:m-1})$
is the probability to measure $\M_m$ given the previous likelihood function $\L_{m-1}=\L(\S\,|\,\M_{1:m-1})$, and
$\L_0(\S)$ is a Bayesian prior.

\subsection{Likelihood function in the limit of weak gradients}

After a single measurement $\M_1$
that yielded a measured angle $\psi_\tau$, 
i.e., $\wt{n}_1 = |\wt{n}_1| e^{i\psi_\tau}$, 
the likelihood function $\L_1=\L(\S|\M_1)$ for $\S=(c,\alpha,\psi)$ reads
\begin{align}
\label{eq:one_measurement}
\L_1 
&= 
\Lambda\,
\exp\bigg( 
	\underbrace{
		\wt{n}_0 |\wt{n}_1| \frac{A}{\nu} 
	}_{\textstyle \kappa_\tau}
	\cos(\psi-\psi_\tau)
\bigg) 
\exp\left(
	-|\wt{n}_1|^2 \frac{\alpha A}{4\nu}
	\cos 2(\psi-\psi_\tau)
\right)\,
\L_0 \quad,
\end{align}
which follows from Eqs.~(\ref{eq:measurement_process}) and (\ref{eq:Bayes_rule}).
Here, 
$\Lambda=\Lambda(c,\alpha,\M_1)$ 
is a prefactor independent of $\psi$, 
see appendix A.3, 
and we used short-hand
$A=2\alpha/(2-\alpha^2)\approx \alpha$ for $\alpha\ll 1$.

Of note,
$\L_1$ contains as a factor a von-Mises distribution for $\psi$, 
$p(\psi)\sim \exp[\kappa_\tau \cos(\psi-\psi_\tau)]$
with measure of concentration
$\kappa_\tau = \wt{n}_0 |\wt{n}_1| A / \nu$, 
see also appendix \ref{app:circular_distributions}.
In the limit of weak concentration gradients, $\alpha_0\ll 1$, this factor dominates
(for likely $\S$ and typical $\M$). 
The second exponential factor in Eq.~(\ref{eq:one_measurement}), 
which results from the off-diagonal entries of the covariance matrix $\Sigma$,  
represents a von-Mises distribution for $2\psi$. 
The corresponding measure of concentration 
$|\wt{n}_1|^2 \alpha A/(4\nu) \sim \alpha_0 \kappa_\tau$ 
is small compared to that of the first factor 
(for likely $\S$ and typical $\M$).
Hence, we can approximate this factor by a constant.

The measure of concentration $\kappa_\tau$ corresponding to a single measurement 
depends on $\M_1$ and is thus itself a random variable.
We can compute the expectation value of $\kappa_\tau^2$ exactly
\begin{align}
\label{eq:kappa_tau}
\langle \kappa_\tau^2 \rangle
&= \alpha_0^2\frac{\langle\wt{n}_0^2|\wt{n}_1|^2\rangle}{\nu_0^2} \\
&= 2\,\SNR + \SNR^2
  + \mathcal{O}\left(\alpha_0^2,\alpha_0^4\nu_0 \right)\quad,
\end{align}
see also appendix A.1. 
For the first moment of $\kappa_\tau$, 
we find an analytic expression in the limit of high signal-to-noise ratio
\begin{equation}
\label{eq:kappa_tau_high_SNR}
\langle \kappa_\tau \rangle \approx \SNR + \frac{1}{2}
\quad \text{ for } 
\SNR \gg 1 \quad, 
\end{equation}
see appendix A.2. 
Accordingly, we can interpret $\langle \kappa_\tau^2 \rangle$ 
as the sum of a squared mean $\langle \kappa_\tau \rangle^2 \approx \SNR^2$,
and a variance 
$\langle \kappa_\tau^2\rangle - \langle \kappa_\tau \rangle^2 \sim \SNR $.

The asymptotic scaling $\langle \kappa_\tau \rangle \sim \SNR$ is consistent with geometric intuition:
the measure of concentration $\kappa_\tau$ is closely related to the circular variance,
which in turn can be estimated by considering
a typical right-angled triangle in the complex plane with small angle $\psi_\tau-\psi_0$ and catheti
$|\mathrm{Re}\, \wt{n}_1 e^{-i\psi_0}| \approx \alpha_0\nu_0/2$ and
$|\mathrm{Im}\, \wt{n}_1 e^{-i\psi_0}| \sim \sqrt{\nu_0}$ for $\SNR\gg 1$.
Hence,
$(2\kappa_\tau)^{-1} \approx 1-\langle \cos(\psi_\tau-\psi_0) \rangle \sim [\sqrt{\nu_0} / (\alpha_0\nu_0/2)]^2 \sim \SNR^{-1}$.

Next, we give an explicit approximation for $\L_1$.
For simplicity, 
we consider the special case, where the Bayesian prior $\L_0(\S)$
is itself a von-Mises distribution in $\psi$,  
centered at $\psi_0$ with measure of concentration $\kappa_0$, 
and the agent possesses perfect knowledge of the other two environmental variables, $c_0$ and $\alpha_0$, 
$\L_0 \sim \exp[\kappa_0 \cos(\psi-\psi_0)]\, \delta(c-c_0) \delta(\alpha-\alpha_0)$. 
This is not a severe restriction, at least not for the absolute concentration $c$,
as agents can estimate $c_0$ very precisely for $\nu_0\gg 1$.
Now, the updated likelihood distribution $\L_1$ is again a von-Mises distribution 
with new measure of concentration $\kappa_1$ and
maximum-likelihood angle $\wh{\psi}_1$, see also appendix \ref{app:circular_distributions}
\begin{equation}
\label{eq:wh_psi}
\kappa_1\, e^{i\wh{\psi}_1} = \kappa_0\, e^{i\psi_0}+\kappa_\tau\, e^{i\psi_\tau} \quad .
\end{equation}
As expected, $\wh{\psi}_1$ is a weighted circular mean of the prior $\psi_0$ and the measured $\psi_\tau$

For the measure of concentration $\kappa_1$ of the updated likelihood function $\L_1$, 
we find from Eq.~(\ref{eq:wh_psi}) with 
$\kappa_\tau \approx \alpha_0 \wt{n}_0|\wt{n}_1|/\nu_0$ 
\begin{align}
\langle \kappa_1^2 \rangle
&= 
\kappa_0^2 +
\langle\kappa_\tau^2\rangle +
\frac{\alpha_0}{\nu_0} \kappa_0 e^{i\psi_0} \langle\wt{n}_0\wt{n}_1\rangle^\ast + \mathrm{c.c.} \notag \\
&= 
\kappa_0^2 + 
2 (1+\kappa_0)\,\SNR + 
\SNR^2 + \mathcal{O}\left(\alpha_0^2,\alpha_0^4\nu_0 \right) \quad.
\label{eq:kappa_1}
\end{align}

\subsection{Sequential estimates in the absence of rotational diffusion}
\label{sec:kappa_m}

We are interested in the marginal likelihood distribution $\L_m(\psi)$
of the estimated gradient angle $\psi$ after $m$ subsequent measurements.
By applying Bayes' update rule Eq.~({\ref{eq:Bayes_rule}) iteratively $m$ times, 
we obtain an approximate formula for $\L_m(\psi)$
as a von-Mises distribution 
\begin{equation}
\L_m(\psi) \sim \exp \left[ \kappa_m \cos(\psi-\wh{\psi}_m) \right] \quad.
\end{equation}
Next, we compute the measure of concentration $\kappa_m$.
In the absence of rotational diffusion, 
subsequent measurements are independent random variables.
This allows us to compute the second moment of $\kappa_m$ 
analogous to Eqs.~(\ref{eq:wh_psi_m}) and (\ref{eq:kappa_1}), 
see appendix A.4 
\begin{align}
\label{eq:kappa_m}
\langle \kappa_m^2 \rangle 
=
\kappa_0^2 + 
2(1+\kappa_0)\,m\SNR + 
m^2 \SNR^2 + 
\mathcal{O}(\alpha_0^2,\alpha_0^4\nu_0) \quad.
\end{align}
Eq.~(\ref{eq:kappa_m}) corroborates how chemotactic agents
increasingly become more confident of their gradient estimates 
as the number $m$ of sequential measurements increases.
Eq.~(\ref{eq:kappa_m}) 
is equivalent to the result for a single long measurement of duration $m\tau$, 
for which the effective signal-to-noise ratio reads $m\SNR$, 
see also appendix \ref{app:no_impatience}.

Asymptotically, 
the root-mean-square expectation value of the measure of concentration,
normalized by the number $m$ of measurements, approaches the signal-to-noise ratio $\SNR$
\begin{align}
\boxed{
\lim _{m\to \infty} \frac{1}{m} \langle\kappa_m^2\rangle^{1/2} = \SNR.
}
\end{align}

\section{Gradient sensing in the presence of rotational diffusion}\label{sec:presence_of_D}

We now consider a chemotactic agent subject to rotational diffusion with $D>0$.
The orientational angle $\true{\psi}(t)$ 
that specifies the direction of the gradient vector $\g$
relative to the material frame of the agent at time $t$, i.e.,\ 
$\cos[\true{\psi}(t)] = \h_1(t)\cdot\g$,
thus becomes a stochastic process.
For simplicity, we consider discrete time dynamics, 
where rotational diffusion events occur at discrete times $t_j=j\tau$.
Thus, the orientation angle $\true{\psi}(t)$ is constant during each interval $\T_j=(t_{j-1},t_j)$
with $\true{\psi}(t)=\psi_j$ for $t\in\T_j$,
with independent random increments $\Delta_j = \psi_{j+1} - \psi_j$,
normally distributed with $\langle \Delta_i\Delta_j\rangle = 2D\tau\,\delta_{ij}$, 
see Fig.~\ref{figure1}B.

\subsection{Rotational diffusion jeopardizes gradient measurements if agents are unaware of it}

We calculate the expected measure of concentration of sequential gradient estimates for $D>0$,
following the calculation for the special case $D=0$ in section \ref{sec:kappa_m}.
For simplicity, we again assume a Bayesian prior of the form
$\L_0(c,\alpha,\psi)\sim \exp[\kappa_0(\psi- \psi_0 )]\,\delta(c-c_0)\,\delta(\alpha-\alpha_0)$.
For agents with rotational diffusion,
subsequent measurements $\M_j$ and $\M_k$ are not independent random variables
because they depend on the underlying stochastic process $\true{\psi}(t)$.
Specifically,
\begin{equation}
\label{eq:n1j_n1k}
\langle \wt{n}_1^{(j)} \rangle = \frac{\alpha_0\nu_0}{2} e^{i\psi_0} \,e^{ - D j\tau }
\text{ and }
\langle \wt{n}_1^{(j)} \wt{n}_1^{(k)}{}^\ast \rangle = 
\frac{ (\alpha_0 \nu_0)^2 }{4} \, e^{-D|k-j|\tau} \quad.
\end{equation}
This correlation marks a crucial difference to the case $D=0$ treated above in Eq.~(\ref{eq:kappa_m}),
where we exploited that subsequent measurements are independent.
For $D>0$, Eq.~(\ref{eq:kappa_m_calc_a}) in appendix A.4 
still holds, but Eq.~(\ref{eq:kappa_m_calc_b}) does not.

We now only use the approximation that $\wt{n}_0^{(j)}$ and $\wt{n}_1^{(j)}$ 
are approximately independent for each measurement,
but account for the correlation of subsequent measurements in Eq.~(\ref{eq:n1j_n1k}).
With this approximation, we obtain
\begin{equation}
\label{eq:kappa_m_D}
\langle \kappa_m^2 \rangle 
= 
\kappa_0^2 + 
\alpha_0^2 \left[
    \sum_j\langle |\wt{n}_1^{(j)}|^2 \rangle + 
    \sum_{j\neq k} \langle \wt{n}_1^{(j)}\wt{n}_1^{(k)}{}^\ast \rangle 
\right] +
\alpha_0\kappa_0 e^{ i\psi_0 } \sum_j \langle\wt{n}_1^{(j)} \rangle^\ast 
+ \mathrm{c.c.}\quad .  
\end{equation}
We compute the sums of expectation values in Eq.~(\ref{eq:kappa_m_D})
by evaluating a (double) geometric series using Eq.~(\ref{eq:n1j_n1k}), see appendix A.4. 
As result, we find
\begin{equation}
\label{eq:kappa_m_D2}
\langle \kappa_m ^2\rangle
= 
\kappa_0^2 + 
2 \Phi_1(m,\kappa_0)\, \SNR +
2 \Phi_2(m)\, \SNR^2 +
\O(\alpha_0^2,\alpha_0^4\nu_0)
\end{equation}
with
$
\Phi_1(m,\kappa_0) = 
m + \kappa_0 \frac{1-e^{- m D\tau}}{e^{D\tau}-1} 
$
and
$
\Phi_2(m) = 
m \frac{1}{e^{D\tau}-1} -     
	e^{D\tau} 	
		\frac{1-e^{- m D\tau}}{\left(e^{D\tau}-1 \right)^2} + 
	\frac{m}{2} 
$.
In the limit of slow diffusion, $D\tau\ll 1$,
we find to leading order in $D\tau$,
$\Phi_1 \approx (1+\kappa_0)m$ for $m\tau\ll D^{-1}$ and
$\Phi_1 \approx m+\kappa_0/(D\tau)$ for $m\tau\gg D^{-1}$, as well as
$\Phi_2 \approx m^2/2$ for $m\tau\ll D^{-1}$ and
$\Phi_2 \approx m/(D\tau)$ for $m\tau\gg D^{-1}$.
This provides an asymptotic scaling for $\kappa_m$,
valid in the limit $\SNR\gg 1$ and slow diffusion, $D\tau\ll 1$
\begin{equation}
\label{eq:kappa_m3}
\boxed{
\langle \kappa_m^2 \rangle^{1/2} \approx 
\begin{cases}
	m\, \SNR & \text{ for } m\tau \ll D^{-1} \\	 
	\sqrt{ \frac{2 m \tau}{D} }\, \SNR/\tau & \text{ for }  m\tau \gg D^{-1}
\end{cases}\quad.
}
\end{equation}
For $D>0$, $\kappa_m$ will initially increase linearly with $m$,
and cross-over to the asymptotic scaling $\kappa_m\sim m^{-1/2}$
beyond a characteristic measurement time $t=m\tau$ on the order of $D^{-1}$.
In fact, 
the condition $\SNR\gg 1$ is not needed for this asymptotic scaling,
provided $D\ll \SNR/\tau$ and $m \gg \SNR^{-1}$.  
(The first condition ensures $\Phi_2\gg \Phi_1$, 
while the second condition implies that the contribution of the Bayesian prior is negligible.)
Note that in the continuum limit $\tau\rightarrow 0$ with $t=m\tau$ fixed,
Eq.~(\ref{eq:kappa_m3}) becomes 
$\langle\kappa(t)^2\rangle ^{1/2} = \sqrt{ 2  t/D} \, \SNR/\tau$, 
where $\SNR/\tau=\alpha_0^2 \lambda c_0 / 2$ is independent of $\tau$.

Eq.~(\ref{eq:kappa_m3}) shows how
an agent subject to rotational diffusion 
that does not take into account its own stochastic motion in its update process 
will erroneously believe that its gradient direction estimate becomes increasingly more accurate 
if measurement time $t=m\tau$ is increased. Yet, this is wrong.

In fact, in an \textit{ensemble} of agents, the estimation errors 
$\delta_j=\wh{\psi}_j-\psi_j$
will eventually become completely randomized.
To illustrate this behavior, 
we characterize the distribution of estimation errors $\delta$ within an \textit{ensemble} of agents,
approximating it by a wrapped normal distribution with variance parameter $\sigma^2_m$.
We show that $\sigma_m^2$ increases as a function of time $t_m$.
For the estimation error of an individual agent, 
we have an approximate iteration rule,
valid for early times, $m\tau\ll D^{-1}$, and high signal-to-noise ratio, $\SNR\gg 1$,
which expresses the new error as an affine interpolation of the previous error and the error of the last measurement
\begin{equation}
\label{eq:maxlik}
\delta_m 
\approx
\frac{{\sigma}^2_\tau}{\wh{\sigma}^2_{m-1}+{\sigma}^2_\tau}
\underbrace{
	\left( \delta_{m-1} - \Delta_{m-1} \right)
}_{
\begin{array}{c} 
\text{\scriptsize previous estimate} \\[-1mm] 
\text{\scriptsize plus noise} 
\end{array}
}
+
\frac{\wh{\sigma}^2_{m-1}}{\wh{\sigma}^2_{m-1}+{\sigma}^2_\tau}
\underbrace{
\left( \arg \wt{n}_1^{(m)} - \psi_m \right)
}_{
\begin{array}{c}
\text{\scriptsize new measurement}
\end{array}}
\quad .
\end{equation}
Here, 
we introduced the variance parameters 
${\sigma}^2_\tau$ and $\wh{\sigma}^2_m$
of wrapped normal distributions approximating von-Mises distributions with
measures of concentration $\kappa_\tau$ and $\kappa_m$ computed above in Eqs.~(\ref{eq:kappa_tau_high_SNR}) and (\ref{eq:kappa_m}), 
such that respective distributions have the same circular variance.
Mathematically, ${\sigma}^2 = - 2\ln I_1(\kappa)/I_0(\kappa)$, see appendix \ref{app:circular_distributions}.
From Eq.~(\ref{eq:maxlik}), we obtain an approximate iteration rule for $\sigma_m^2$
\begin{align}
\label{eq:iteration_maxilik}
\sigma_m^2 
&\approx
\left( \frac{{\sigma}^2_\tau}{\wh{\sigma}^2_{m-1}+{\sigma}^2_\tau} \right)^2
	\left( \sigma^2_{m-1} + 2D\tau \right) + 
\left( \frac{\wh{\sigma}^2_{m-1}}{\wh{\sigma}^2_{m-1}+{\sigma}^2_\tau} \right)^2
{\sigma}^2_\tau \quad.
\end{align}
This expression suggests that $\sigma_m^2$ grows asymptotically as $\sqrt{2D\tau\,m}$, see appendix A.5. 
Correspondingly, the circular variance 
$\CV = 1-e^{-\sigma^2/2}$ 
of the distribution $p(\delta)$ should increase, eventually
converging to $1$ for $m\tau \gg D^{-1}$.
Although the specific assumptions made in the derivation of Eq.~(\ref{eq:iteration_maxilik}) do not hold in this limit,
simulations corroborate this simple picture, see Fig.~\ref{figure2}A.

In conclusion, agents not aware of their own rotational diffusion, 
will arrive at erroneous gradient estimates.
The reason is that past measurements will have become partially invalidated by rotational diffusion, 
yet are nonetheless incorporated in the gradient estimates with full weight.
Concomitantly, 
the precision that \textit{individual} agents estimate for their own gradient measurement 
does not reflect the true accuracy, i.e., 
the dispersion of maximum-likelihood estimates in an \textit{ensemble} of agents.
Individual agents are `over-confident' of their own estimates.

\begin{figure}
\includegraphics[width=0.8\textwidth]{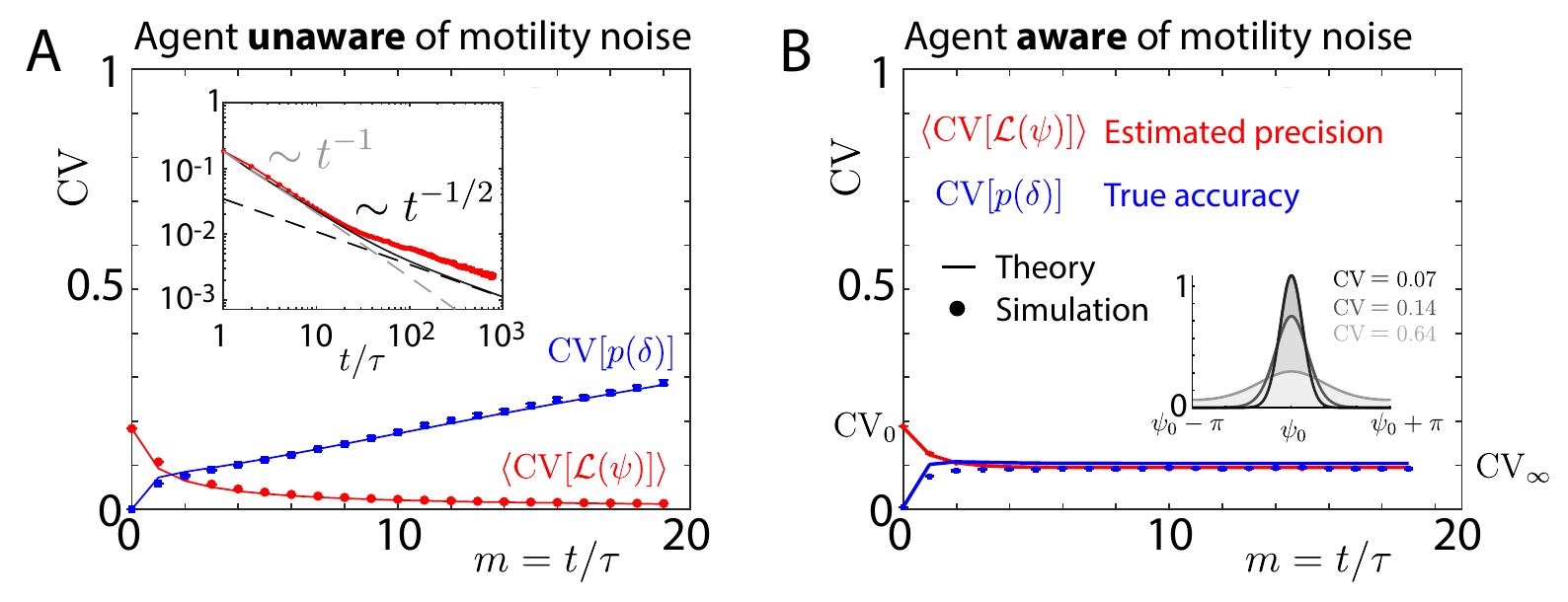}
\caption[]{
\textbf{Estimated precision and true accuracy of Bayesian gradient sensing.}
\textbf{A.}
\textit{Agents unaware of own rotational diffusion.}
Each individual agents computes a likelihood distribution $\L_m(\psi)$ at each time step, 
with maximum-likelihood direction angle $\wh{\psi}_m$ and circular variance $\CV[\L_m(\psi)]$.
Shown is the ensemble-averaged circular variance $\langle\CV[\L_m(\psi)]\rangle$ (\textit{estimated precision}, red), and 
the circular variance $\CV[p(\delta_m)]$ of estimation errors $\delta_m=\wh{\psi}_m-\psi_m$ within the ensemble of agents
(\textit{true accuracy}, blue). 
Solid lines represent the analytical results, Eq.~(\ref{eq:kappa_m_D2}) and (\ref{eq:iteration_maxilik}),
for estimated precision and accuracy, respectively.
The accuracy converges to one, corresponding to the randomization of estimated angles $\wh{\psi}$.
At the same time, 
the estimated precision converges to zero, 
displaying a cross-over between two scaling regimes as predicted by Eq.~(\ref{eq:kappa_m3}), see inset.
\textbf{B.}
\textit{Agents aware of own rotational diffusion.}
Same as panel A, but agents take into account their rotational diffusion in a prediction step for $\L(\psi)$. 
Solid lines represent the analytical results, Eq.~(\ref{eq:kappa_m_discounted}) and (\ref{eq:iteration_maxilik}).
Estimated precision and accuracy converge to the same limit value, $\CV_\infty$. 
Inset illustrates circular distributions with circular variance 
$0.07$ (black), $0.14$ (gray), $0.64$ (light-gray), 
using von-Mises distributions centered at an arbitrary $\psi_0$.
Error bars represent s.e.m.\ (determined by bootstrapping for an ensemble of $n=5000$ agents, occasionally smaller than symbols).
Parameters: $\nu_0=5000$, $\alpha_0=0.03$, $D\,\tau=0.05$; 
Bayesian prior: $\kappa_0=3.09 \approx \langle\kappa_\tau^2\rangle^{1/2}$, $\psi_0=0$. 
To make analytical results comparable to simulated circular variances, we used the formula
$\CV=1-I_1(\kappa)/I_0(\kappa)$ 
to convert the measure of concentration $\kappa$ of the von-Mises distributions 
in Eq.~(\ref{eq:kappa_m_D2}) [panel (a)] and Eq.~(\ref{eq:kappa_m_discounted}) [panel (b)]
to a circular variance. 
Similarly, we used
$\CV=1-\exp(-\sigma^2/2)$
to convert the variance paremeter $\sigma^2$ of the wrapped normal distribution 
in Eq.~(\ref{eq:iteration_maxilik}) 
to a circular variance, 
where we additionally used
$\wh{\sigma}^2 = - 2\ln I_1(\kappa)/I_0(\kappa)$ 
to relate $\kappa_m$ from Eq.~(\ref{eq:kappa_m_D2}) [panel (a)] and Eq.~(\ref{eq:kappa_m_discounted}) [panel (b)]
to $\wh{\sigma}^2_m$ in Eq.~(\ref{eq:iteration_maxilik}).
}		
\label{figure2}
\end{figure}

\subsection{Agents aware of own rotational diffusion}
\label{sec:agents_aware}

We now consider an agent that correctly takes into account its own rotational diffusion 
before updating the likelihood distribution $\L(\S)$ of estimated concentration gradients. 
Following the terminology of the known Kalman filter algorithm, 
we consider in addition to the \textit{update step}, Eq.~(\ref{eq:Bayes_rule}),
which describes the incorporation of new measurement information, 
an additional \textit{prediction step}
that describes the change of $\L(\S)$
due to rotational diffusion, see Fig.~\ref{figure1}C.
In its most general form, this prediction step 
is given by a Chapman-Kolmogorov equation
\begin{align}
\label{eq:prediction}
\L'_{m-1} = \L'(\S'|\M _{1:m-1}) = 
\int_\S P(\S'|\S)\,\L(\S|\M _{1:m-1})\,d\S \quad .
\end{align}
Here, $P(\S'|\S)$ is the transition probability 
from state $\S$ to state $\S'$ at time $t_{m-1}$.
In our case,
$P(\psi'|\psi)$
is a wrapped normal distribution with zero mean and variance $2D\tau$, 
while the other two state variables do not change, i.e.,
$P(\S'|\S) = P(\psi'|\psi)\,\delta(c'-c)\,\delta(\alpha'-\alpha)$ is independent of time $t$.

To make analytical progress, we again assume perfect knowledge of concentration $c_0$ and gradient strength $\alpha_0$,
i.e., a Bayesian prior of the form $\L_0(\S) \sim \exp[\kappa_0 \cos(\psi-\psi_0)]\,\delta(c-c_0)\,\delta(\alpha-\alpha_0)$.
In the limit of high signal-to-noise ratio, $\SNR\gg 1$, and slow diffusion, $D\tau \ll 1$, 
we can approximate all factors in Eqs.~(\ref{eq:Bayes_rule}) and (\ref{eq:prediction}) 
by von-Mises distributions with appropriate measures of concentrations.
In this limit, the update step corresponds to the (normalized) product of two von-Mises distributions, 
while the prediction step corresponds to the convolution of two such distributions.
From the calculus of directional distributions, see appendix \ref{app:circular_distributions},
we obtain a recursive relation for the measure of concentration $\kappa_m$ 
of a von-Mises distribution approximating $\L_m(\psi)$
\begin{equation}
\label{eq:kappa_m_discounted}
\boxed{
\kappa_m = 
    \left( \frac{1}{ \kappa_{m-1} } + 2D\tau \right)^{-1} 
    + \kappa_{\tau} \quad .      
} 
\end{equation}
Here, we assume $\kappa_m \approx \langle \kappa_m\rangle$, $\kappa_\tau \approx \langle \kappa_\tau\rangle$, 
which is valid for $\SNR\gg 1$.
For completeness, we list all assumptions made in deriving Eq.~(\ref{eq:kappa_m_discounted}):
(i) high molecule count, $\nu_0\gg 1$ (enabling the diffusion approximation in the measurement model Eq.~(\ref{eq:measurement_process})), 
(ii) weak gradient, $\alpha_0\ll 1$ (allowing us to approximate likelihood distributions by von-Mises distributions),
(iii) high signal-to-noise ratio, $\SNR\gg 1$ (allowing us to equate the precision estimated by an agent with its expectation value),
(iv) slow diffusion, $D\tau \ll 1$ 
(which, together with $\SNR\gg 1$, ensures 
that the simple formulas Eq.~(\ref{eq:VM_conv}) and Eq.~(\ref{eq:VM_mult}) 
can be used for the measure of concentration of convolutions and normalized products of von-Mises distributions, respectively).

\subsection{Theoretical limit of gradient sensing precision in the presence of rotational diffusion}

The iteration rule Eq.~(\ref{eq:kappa_m_discounted}) 
for the measure of concentration $\kappa_m$ after $m$ measurements
defines a monotonically increasing sequence
with limit value $\kappa_\infty$
(given by a root of the quadratic equation $\kappa_\infty(\kappa_\infty+\kappa_\tau) = \kappa_\tau/(D\tau)$).
%
Provided measurement intervals are short, $\tau\ll (D\,\SNR/\tau)^{-1/2}$, we have
\begin{equation}
\label{eq:k_infty}
\boxed{
\kappa_\infty 
\approx \sqrt{ \frac{\kappa_\tau}{2D\tau} } 
= \sqrt{\frac{\lambda}{4\,c_0 D}} \, |\nabla c| a  \quad.
} 
\end{equation}
This result for $\kappa_\infty$ highlights the competition between 
\textit{information gain} with rate $\kappa_\tau/\tau \approx \SNR/\tau$ and
\textit{information loss} by rotational diffusion with rotational diffusion coefficient $D$.
Mathematically, Eq.~(\ref{eq:k_infty}) is valid in a `sandwiched' limit
$(\SNR/\tau)^{-1} \ll \tau \ll (D\,\SNR/\tau)^{-1/2}$.
Note that in the opposite limit, $\tau \gg (D\,\SNR/\tau)^{-1/2}$, 
each rotational diffusion event would erase all previous measurements.

We can understand the scaling for the limit value in Eq.~(\ref{eq:k_infty}) intuitively as follows.
We expect $\kappa_m$ to increase linearly with total measurement time $t=m\tau$ for $t\ll t_\infty$, 
and to saturate to a limit value $\kappa_\infty$ for $t\gg t_\infty$,
where the cross-over time $t_\infty=t_\infty(D)$ is a yet unknown function of the rotational diffusion coefficient $D$.
Such a saturation curve suggests a scaling relation, 
$ t_\infty/\tau \sim k_\infty/k_\tau $.
Any measurements taken at a time $t$ in the past 
will have been corrupted by rotational diffusion to an extent
that they do not serve to increase the measure of concentration above a value 
$(D t)^{-1}$.
Thus, the cross-over time $t_\infty$ must satisfy
$D t_\infty \sim k_\infty^{-1}$.
We conclude 
$t_\infty^2 \sim \tau / (D \kappa_\tau)$, 
hence 
$\kappa_\infty \sim \sqrt{ \kappa_\tau / (D\tau) }$.

With Eq.~(\ref{eq:k_infty}), we can characterize the distribution of
estimation errors $\delta_m=\wh{\psi}_m-\psi_m$ within an ensemble of agents. 
By Eq.~(\ref{eq:iteration_maxilik}), the variance parameter of this distribution converges to
$\sigma^2_\infty = \lim_{m\rightarrow\infty} \sigma_m^2 \approx \kappa_\infty^{-1}$.
Thus, 
$\sigma_\infty^2 \approx \wh{\sigma}_\infty^2$, i.e., 
the \textit{estimated precision}
$\wh{\sigma}^2_\infty \approx \kappa_\infty^{-1} = [2D\tau/\SNR]^{1/2}$} 
of gradient sensing of each individual agent
is a faithful estimator for the \textit{true accuracy}, i.e., the
dispersion $\sigma^2_\infty$ of maximum-likelihood estimates within an ensemble, 
provided the agents know their rotational diffusion coefficient $D$.
Fig.~\ref{figure2}B corroborates this finding for the equivalent measure of circular variance.

More generally, we can consider agents that assume a value $\wh{D}$ for their rotational diffusion coefficient
when performing their prediction step Eq.~(\ref{eq:prediction}), 
while the true rotational diffusion coefficient is $D$. 
In this case, the variance parameter $\sigma_\infty^2$ of the distribution $p(\delta)$ of estimation errors 
follows from Eqs.~(\ref{eq:iteration_maxilik}) and (\ref{eq:kappa_m_discounted}) in the long-time limit as
\begin{equation}
\boxed{
\sigma^2_\infty = \sqrt{\frac{2\,D\tau}{\SNR}} \cdot \frac{D+\wh{D}}{\sqrt{4D\wh{D}}} \quad,
}
\end{equation}
which attains the minimal value $\sigma^2_\infty=\wh{\sigma}^2_\infty$ exactly for $\wh{D}=D$.

\section{Discussion}

\paragraph*{Summary of results.}
We considered a minimal model of gradient sensing in the presence of both sensing and motility noise.
We derived analytical results for sequential Bayesian estimation
by chemotactic agents that undergo rotational diffusion.
Gradient sensing fails
if agents are not aware of their own rotational diffusion, 
because agents extend temporal averaging infinitely into the past. 
Concomitantly, the estimated gradient direction decorrelates from the true direction on the time-scale of rotational diffusion, 
while agents erroneously believe that their estimates become more and more precise as function of total measurement time $t$.
Interestinlgy, we find an abnormal asymptotic scaling of the estimated variance of gradient estimates,
$\CV[\L(\psi)] \sim t^{-1/2}$, 
see Eq.~(\ref{eq:kappa_m3}).
This abnormal scaling is a signature of erroneous state estimation 
and is intimately related to the properties of circular statistics.
This signature could be tested for in real-world applications, e.g., of bearing tracking.

In contrast, if agents know their own diffusion coefficient, 
sequential Bayesian estimation yields accurate estimates of gradient direction.
These estimates are both \textit{faithful} and \textit{self-consistent}, 
i.e., estimation errors are unbiased and 
individual agents can estimate how large their estimation error is in each time step.
In fact, the estimated precision that individual agents assign to their individual direction estimate
converges to the true accuracy, i.e., the dispersion of maximum-likelihood estimates within an ensemble of agents.
Remarkably, the ultimate precision of gradient sensing scales with the square root of the rotational diffusion coefficient $D$ as
$\CV[\L(\psi)] \sim \sqrt{D}$.
Intuitively, agents extend temporal averaging only over the recent past defined by a time span of duration $t_\infty\sim D^{-1/2}$.
Measurements taken before this time span
still contain partial information on the current gradient direction
as long as they do not extend beyond the rotational diffusion time $D^{-1}$. 
Yet, these measurements are already corrupted too much by rotational diffusion as that they could 
add to the precision achieved by temporal averaging only over the time span $t_\infty$.
In fact, these past measurement would make the estimate worse if they were included.
Mathematically, this reflects a difference between estimating a vectorial quantity, e.g., gradient direction, 
as opposed to estimating a scalar quantity, see appendix \ref{app:filter}.

\paragraph*{Previous theoretical limits.}
Our result on the optimal time span for temporal averaging 
is different from previous work \cite{Strong1998},
which suggested that temporal comparison should be extended over a time span set by the rotational diffusion time, 
i.e., $t_\infty \sim D^{-1}$.
This previous work addressed a different sensing and chemotaxis strategy
based on \textit{temporal comparison}, which measures a scalar quantity. 
It turns out that this makes a crucial difference.
In short, temporal comparsion relies on active motion of chemotactic agents within a spatial concentration gradient,
such that concentration differences in space become encoded in a temporal change of local concentration measurements.
Thereby, the cells estimate only a scalar quantity, 
the component of the concentration gradient along their direction of motion ($\v\cdot\nabla c$).
If a cell's swimming direction $\v/|\v|$ decorrelates on a time-scale $D^{-1}$ due to rotational diffusion, 
an optimal filter should indeed discount previous measurements on the same time-scale, 
i.e., downweight measurements taken a time $\Delta t$ before by a factor $\sim\exp(-D\,\Delta t)$.
The strategy of temporal comparison is different from \textit{spatial comparison} as considered here,
where chemotactic agents estimate concentration gradients by comparing concentrations across their diameter.
Bacteria performing run-and-tumble chemotaxis employ temporal comparison \cite{Segall1986}, 
while most eukaryotic cells with crawling motility employ spatial comparison \cite{Zigmond1977,Devreotes1988,Arkowitz1999,Manahan2004,Sarris2015}.
A third chemotaxis strategy is represented by marine sperm cells navigating along helical paths \cite{Alvarez2014}.
Although these cells effectively use only a single sensor, 
their chemotaxis can be mapped on the case of spatial comparison considered here:
while moving along helical paths, these cells `visit' different sensor positions during one helical turn.

In conclusion, our work identified a crucial difference 
regarding the optimal time span of temporal averaging
between the chemotaxis strategies of temporal and spatial comparison
if both sensing and motility noise are present. 

\paragraph*{Typical parameters.}
Typical rotational diffusion coefficients for the bacterium \textit{E. coli} are $D\sim 0.1\,\mathrm{s}^{-1}$, 
close to the theoretical lower limit of a passive particle of same size and shape. 
For ten-fold larger sperm cells, active fluctuations dominate \cite{Ma2014}, resulting in an estimate $D\sim 0.01-0.1\,\mathrm{s}^{-1}$ \cite{Friedrich2008}.
The motility of crawling \textit{Dictyostelium} cells was characterized by a persistence time of $\sim 5\, \mathrm{min}$
in the absence of chemoactractant \cite{vanHaastert2010}, 
which sets an effective rotational diffusion coefficient $D\sim 0.003\,\mathrm{s}^{-1}$. 
For immune T cells in three-dimensional tissue, 
a persistence time of $\sim 1\,\mathrm{min}$ was found (displaying a characteristic speed dependence) \cite{Jerison2019}.
If binding of signaling molecules to receptors on the cell surface is diffusion-limited, 
we can estimate the rate of binding by
$\lambda c_0=4\pi D_c\, a c_0$ for a perfectly absorbing spherical cell of radius $a$, 
where $D_c$ denotes the translational diffusion coefficient of signaling molecules \cite{Berg1993}. 
For typical values ($D_c\sim 300\,\micron^2\,\mathrm{s}^{-1}$, $a\sim 10\,\micron$, $c_0\sim 1\,\mathrm{nM}$), we estimate
$\lambda c_0 \sim 10^4\,\mathrm{s}^{-1}$.
Thus, 
for a concentration gradient of either 
$\alpha_0=1\%$, $0.5\%$, or $0.1\%$ across the diameter of a cell, 
and a measurement time $\tau=10\,\mathrm{s}$,
we estimate signal-to-noise ratios of gradient sensing, 
$\SNR\approx 3.5$, $\approx 0.9$, and $\approx 0.03$, respectively.
Assuming $D\sim 0.003\,\mathrm{s}^{-1}$ \cite{vanHaastert2010},
our main result, Eq.~(\ref{eq:k_infty}), predicts for the ultimate precision of gradient sensing
$\CV_\infty \approx 0.07$, $\approx 0.14$, and $\approx 0.64$
for these three gradient strengths, respectively
(see inset of Fig.~\ref{figure2}B for visualization).
Reversible binding of signaling molecules effectively increases sensing noise, 
thus decreasing the signal-to-noise ratio by a constant prefactor \cite{Bialek2005,Endres2008,Kaizu2014}.
Some cells, such as sperm cells, respond chemotacticly even at pico-molar concentrations \cite{Strunker2015}, 
corresponding to respectively lower signal-to-noise ratios \cite{Kashikar2012,Kromer2018}.

Bayesian estimation sets a lower bound for the precision of gradient sensing.
This theoretical limit is relevant at high noise levels, 
but may be less so if noise is low.
For the slime mold \textit{Dictyostelium}, it was shown that at signal-to-noise ratios ($\SNR$) below one, 
the efficacy of chemotaxis was well characterized by $\SNR$ alone,
while noise of downstream intracellular signaling \cite{Ueda2007} 
becomes also relevant at high $\SNR$ \cite{Endres2008,Fuller2010,Amselem2012}.
While several of our analytical results were derived for high $\SNR$, 
scaling relations persist for low $\SNR$ and are confirmed by numerical simulations. 

\paragraph*{Biochemical implementation.}
Storing the likelihood distribution of estimated gradient directions, or just a proxy thereof, requires internal memory. 
We speculate that the distribution of chemotactic effector molecules on the cell boundary 
as considered in recent models such as LEGI (\textit{local excitation, global inhibition}) \cite{Ma2004}, 
or balanced-inactivation models \cite{Levine2006} can indeed serve as a such a proxy.
While the position of a concentration peak in such a distribution can represent a maximum likelihood estimate, 
its amplitude could encode a level of certainty.
Similarly, the directional persistence and polarization of crawling cells represents a form of effective memory \cite{Skoge2014,Andrews2007}. 
For marine sperm cells,
the axis of their helical paths likewise represents a consolidated memory of previous noisy concentration measurements \cite{Friedrich2009}.

\paragraph*{Time-varying gradients.}
Here, we considered only static concentration gradients.
Yet, the general framework developed here generalizes in a straight-forward manner to time-varying environments, 
provided their temporal statistics is known to the agent.
For example, analogous results apply if instead of rotational diffusion of the chemotactic agent, 
it is the direction of the concentration gradient that changes stochastically with a correlation time $D^{-1}$
(e.g., due to an explicit time dependence of concentration fields, or 
due to active motion of the agent within a spatially complex concentration field).
The seminal infotaxis strategy proposed a Bayesian framework for navigation in time-dependent concentration fields, 
using time-averaged properties of scalar turbulence \cite{Vergassola2007}.
In the presence of motility noise, this problem becomes considerably harder, 
for which our work can serve as a first step.

\begin{acknowledgments}
MN and BMF are supported by the German National Science Foundation (DFG) 
through the Excellence Initiative by the German Federal and State Governments 
(Clusters of Excellence cfaed EXC-1056 and PoL EXC-2068),
as well as DFG grant FR3429/3-1 to BMF.
We thank all members of the `Biological Algorithms Group' for stimulating discussions.
\end{acknowledgments}

\bibliographystyle{plain}
\bibliography{bayesian_gradient_sensing}

\begin{thebibliography}{57}
\expandafter\ifx\csname natexlab\endcsname\relax\def\natexlab#1{#1}\fi
\expandafter\ifx\csname bibnamefont\endcsname\relax
  \def\bibnamefont#1{#1}\fi
\expandafter\ifx\csname bibfnamefont\endcsname\relax
  \def\bibfnamefont#1{#1}\fi
\expandafter\ifx\csname citenamefont\endcsname\relax
  \def\citenamefont#1{#1}\fi
\expandafter\ifx\csname url\endcsname\relax
  \def\url#1{\texttt{#1}}\fi
\expandafter\ifx\csname urlprefix\endcsname\relax\def\urlprefix{URL }\fi
\providecommand{\bibinfo}[2]{#2}
\providecommand{\eprint}[2][]{\url{#2}}

\bibitem[{\citenamefont{Berg and Brown}(1972)}]{Berg1972}
\bibinfo{author}{\bibfnamefont{H.~C.} \bibnamefont{Berg}} \bibnamefont{and}
  \bibinfo{author}{\bibfnamefont{D.~A.} \bibnamefont{Brown}},
  \bibinfo{journal}{Nature} \textbf{\bibinfo{volume}{239}},
  \bibinfo{pages}{500} (\bibinfo{year}{1972}).

\bibitem[{\citenamefont{Eisenbach and Giojalas}(2006)}]{Eisenbach2006}
\bibinfo{author}{\bibfnamefont{M.}~\bibnamefont{Eisenbach}} \bibnamefont{and}
  \bibinfo{author}{\bibfnamefont{L.}~\bibnamefont{Giojalas}},
  \bibinfo{journal}{Nat. Rev. Mol. Cell Biol.} \textbf{\bibinfo{volume}{7}},
  \bibinfo{pages}{276} (\bibinfo{year}{2006}).

\bibitem[{\citenamefont{Alvarez et~al.}(2014)\citenamefont{Alvarez, Friedrich,
  Gompper, and Kaupp}}]{Alvarez2014}
\bibinfo{author}{\bibfnamefont{L.}~\bibnamefont{Alvarez}},
  \bibinfo{author}{\bibfnamefont{B.~M.} \bibnamefont{Friedrich}},
  \bibinfo{author}{\bibfnamefont{G.}~\bibnamefont{Gompper}}, \bibnamefont{and}
  \bibinfo{author}{\bibfnamefont{U.~B.} \bibnamefont{Kaupp}},
  \bibinfo{journal}{Trends Cell Biol.} \textbf{\bibinfo{volume}{24}},
  \bibinfo{pages}{198} (\bibinfo{year}{2014}).

\bibitem[{\citenamefont{Devreotes and Zigmond}(1988)}]{Devreotes1988}
\bibinfo{author}{\bibfnamefont{P.~N.} \bibnamefont{Devreotes}}
  \bibnamefont{and} \bibinfo{author}{\bibfnamefont{S.~H.}
  \bibnamefont{Zigmond}}, \bibinfo{journal}{Ann. Rev. Cell Biol.}
  \textbf{\bibinfo{volume}{4}}, \bibinfo{pages}{649} (\bibinfo{year}{1988}).

\bibitem[{\citenamefont{Zigmond}(1977)}]{Zigmond1977}
\bibinfo{author}{\bibfnamefont{S.~H.} \bibnamefont{Zigmond}},
  \bibinfo{journal}{J. Cell Biol.} \textbf{\bibinfo{volume}{75}},
  \bibinfo{pages}{606} (\bibinfo{year}{1977}).

\bibitem[{\citenamefont{Gregor et~al.}(2007)\citenamefont{Gregor, Tank,
  Wieschaus, and Bialek}}]{Gregor2007}
\bibinfo{author}{\bibfnamefont{T.}~\bibnamefont{Gregor}},
  \bibinfo{author}{\bibfnamefont{D.~W.} \bibnamefont{Tank}},
  \bibinfo{author}{\bibfnamefont{E.~F.} \bibnamefont{Wieschaus}},
  \bibnamefont{and} \bibinfo{author}{\bibfnamefont{W.}~\bibnamefont{Bialek}},
  \bibinfo{journal}{Cell} \textbf{\bibinfo{volume}{130}}, \bibinfo{pages}{153}
  (\bibinfo{year}{2007}).

\bibitem[{\citenamefont{Berg and Purcell}(1977)}]{Berg1977}
\bibinfo{author}{\bibfnamefont{H.~C.} \bibnamefont{Berg}} \bibnamefont{and}
  \bibinfo{author}{\bibfnamefont{E.~M.} \bibnamefont{Purcell}},
  \bibinfo{journal}{Biophys. J.} \textbf{\bibinfo{volume}{20}},
  \bibinfo{pages}{193} (\bibinfo{year}{1977}).

\bibitem[{\citenamefont{Bialek and Setayeshgar}(2005)}]{Bialek2005}
\bibinfo{author}{\bibfnamefont{W.}~\bibnamefont{Bialek}} \bibnamefont{and}
  \bibinfo{author}{\bibfnamefont{S.}~\bibnamefont{Setayeshgar}},
  \bibinfo{journal}{Proc. Natl. Acad. Sci. U.S.A.}
  \textbf{\bibinfo{volume}{102}}, \bibinfo{pages}{10040}
  (\bibinfo{year}{2005}).

\bibitem[{\citenamefont{Kaizu et~al.}(2014)\citenamefont{Kaizu, {De Ronde},
  Paijmans, Takahashi, Tostevin, and {ten Wolde}}}]{Kaizu2014}
\bibinfo{author}{\bibfnamefont{K.}~\bibnamefont{Kaizu}},
  \bibinfo{author}{\bibfnamefont{W.}~\bibnamefont{{De Ronde}}},
  \bibinfo{author}{\bibfnamefont{J.}~\bibnamefont{Paijmans}},
  \bibinfo{author}{\bibfnamefont{K.}~\bibnamefont{Takahashi}},
  \bibinfo{author}{\bibfnamefont{F.}~\bibnamefont{Tostevin}}, \bibnamefont{and}
  \bibinfo{author}{\bibfnamefont{P.~R.} \bibnamefont{{ten Wolde}}},
  \bibinfo{journal}{Biophys. J.} \textbf{\bibinfo{volume}{106}},
  \bibinfo{pages}{976} (\bibinfo{year}{2014}).

\bibitem[{\citenamefont{Rappel and Levine}(2008)}]{Rappel2008}
\bibinfo{author}{\bibfnamefont{W.~J.} \bibnamefont{Rappel}} \bibnamefont{and}
  \bibinfo{author}{\bibfnamefont{H.}~\bibnamefont{Levine}},
  \bibinfo{journal}{Proc. Natl. Acad. Sci. U.S.A.}
  \textbf{\bibinfo{volume}{105}}, \bibinfo{pages}{19270}
  (\bibinfo{year}{2008}).

\bibitem[{\citenamefont{Hu et~al.}(2010)\citenamefont{Hu, Chen, Rappel, and
  Levine}}]{Hu2010}
\bibinfo{author}{\bibfnamefont{B.}~\bibnamefont{Hu}},
  \bibinfo{author}{\bibfnamefont{W.}~\bibnamefont{Chen}},
  \bibinfo{author}{\bibfnamefont{W.-J.} \bibnamefont{Rappel}},
  \bibnamefont{and} \bibinfo{author}{\bibfnamefont{H.}~\bibnamefont{Levine}},
  \bibinfo{journal}{Phys. Rev. Lett.} \textbf{\bibinfo{volume}{105}},
  \bibinfo{pages}{048104} (\bibinfo{year}{2010}).

\bibitem[{\citenamefont{Endres and Wingreen}(2008)}]{Endres2008}
\bibinfo{author}{\bibfnamefont{R.~G.} \bibnamefont{Endres}} \bibnamefont{and}
  \bibinfo{author}{\bibfnamefont{N.~S.} \bibnamefont{Wingreen}},
  \bibinfo{journal}{Proc. Natl. Acad. Sci. U. S. A}
  \textbf{\bibinfo{volume}{105}}, \bibinfo{pages}{15749}
  (\bibinfo{year}{2008}).

\bibitem[{\citenamefont{{ten Wolde} et~al.}(2016)\citenamefont{{ten Wolde},
  Becker, Ouldridge, and Mugler}}]{tenWolde2016}
\bibinfo{author}{\bibfnamefont{P.~R.} \bibnamefont{{ten Wolde}}},
  \bibinfo{author}{\bibfnamefont{N.~B.} \bibnamefont{Becker}},
  \bibinfo{author}{\bibfnamefont{T.~E.} \bibnamefont{Ouldridge}},
  \bibnamefont{and} \bibinfo{author}{\bibfnamefont{A.}~\bibnamefont{Mugler}},
  \bibinfo{journal}{J. Stat. Phys.} \textbf{\bibinfo{volume}{162}},
  \bibinfo{pages}{1395} (\bibinfo{year}{2016}).

\bibitem[{\citenamefont{Van~Haastert and Postma}(2007)}]{vanHaastert2007}
\bibinfo{author}{\bibfnamefont{P.~J.} \bibnamefont{Van~Haastert}}
  \bibnamefont{and} \bibinfo{author}{\bibfnamefont{M.}~\bibnamefont{Postma}},
  \bibinfo{journal}{Biophys. J.} \textbf{\bibinfo{volume}{93}},
  \bibinfo{pages}{1787} (\bibinfo{year}{2007}).

\bibitem[{\citenamefont{Mortimer et~al.}(2009)\citenamefont{Mortimer, Feldner,
  Vaughan, Vetter, Pujic, Rosoff, Burrage, Dayan, Richards, and
  Goodhill}}]{Mortimer2009}
\bibinfo{author}{\bibfnamefont{D.}~\bibnamefont{Mortimer}},
  \bibinfo{author}{\bibfnamefont{J.}~\bibnamefont{Feldner}},
  \bibinfo{author}{\bibfnamefont{T.}~\bibnamefont{Vaughan}},
  \bibinfo{author}{\bibfnamefont{I.}~\bibnamefont{Vetter}},
  \bibinfo{author}{\bibfnamefont{Z.}~\bibnamefont{Pujic}},
  \bibinfo{author}{\bibfnamefont{W.~J.} \bibnamefont{Rosoff}},
  \bibinfo{author}{\bibfnamefont{K.}~\bibnamefont{Burrage}},
  \bibinfo{author}{\bibfnamefont{P.}~\bibnamefont{Dayan}},
  \bibinfo{author}{\bibfnamefont{L.~J.} \bibnamefont{Richards}},
  \bibnamefont{and} \bibinfo{author}{\bibfnamefont{G.~J.}
  \bibnamefont{Goodhill}}, \bibinfo{journal}{Proc. Natl. Acad. Sci. U.S.A.}
  \textbf{\bibinfo{volume}{106}}, \bibinfo{pages}{10296}
  (\bibinfo{year}{2009}).

\bibitem[{\citenamefont{Fuller et~al.}(2010)\citenamefont{Fuller, Chen, Adler,
  Groisman, Levine, Rappel, and Loomis}}]{Fuller2010}
\bibinfo{author}{\bibfnamefont{D.}~\bibnamefont{Fuller}},
  \bibinfo{author}{\bibfnamefont{W.}~\bibnamefont{Chen}},
  \bibinfo{author}{\bibfnamefont{M.}~\bibnamefont{Adler}},
  \bibinfo{author}{\bibfnamefont{A.}~\bibnamefont{Groisman}},
  \bibinfo{author}{\bibfnamefont{H.}~\bibnamefont{Levine}},
  \bibinfo{author}{\bibfnamefont{W.-J.} \bibnamefont{Rappel}},
  \bibnamefont{and} \bibinfo{author}{\bibfnamefont{W.~F.}
  \bibnamefont{Loomis}}, \bibinfo{journal}{Proc. Natl. Acad. Sci. U.S.A.}
  \textbf{\bibinfo{volume}{107}}, \bibinfo{pages}{9656} (\bibinfo{year}{2010}).

\bibitem[{\citenamefont{Amselem et~al.}(2012)\citenamefont{Amselem, Theves,
  Bae, Beta, and Bodenschatz}}]{Amselem2012}
\bibinfo{author}{\bibfnamefont{G.}~\bibnamefont{Amselem}},
  \bibinfo{author}{\bibfnamefont{M.}~\bibnamefont{Theves}},
  \bibinfo{author}{\bibfnamefont{A.}~\bibnamefont{Bae}},
  \bibinfo{author}{\bibfnamefont{C.}~\bibnamefont{Beta}}, \bibnamefont{and}
  \bibinfo{author}{\bibfnamefont{E.}~\bibnamefont{Bodenschatz}},
  \bibinfo{journal}{Phys. Rev. Lett.} \textbf{\bibinfo{volume}{109}},
  \bibinfo{pages}{1} (\bibinfo{year}{2012}).

\bibitem[{\citenamefont{Brumley et~al.}(2019)\citenamefont{Brumley, Carrara,
  Hein, Yawata, Levin, and Stocker}}]{Brumley2019}
\bibinfo{author}{\bibfnamefont{D.~R.} \bibnamefont{Brumley}},
  \bibinfo{author}{\bibfnamefont{F.}~\bibnamefont{Carrara}},
  \bibinfo{author}{\bibfnamefont{A.~M.} \bibnamefont{Hein}},
  \bibinfo{author}{\bibfnamefont{Y.}~\bibnamefont{Yawata}},
  \bibinfo{author}{\bibfnamefont{S.~A.} \bibnamefont{Levin}}, \bibnamefont{and}
  \bibinfo{author}{\bibfnamefont{R.}~\bibnamefont{Stocker}},
  \bibinfo{journal}{Proc. Natl. Acad. Sci. U.S.A.}
  \textbf{\bibinfo{volume}{166}}, \bibinfo{pages}{10792}
  (\bibinfo{year}{2019}).

\bibitem[{\citenamefont{Segall et~al.}(1986)\citenamefont{Segall, Block, and
  Berg}}]{Segall1986}
\bibinfo{author}{\bibfnamefont{J.~E.} \bibnamefont{Segall}},
  \bibinfo{author}{\bibfnamefont{S.~M.} \bibnamefont{Block}}, \bibnamefont{and}
  \bibinfo{author}{\bibfnamefont{H.~C.} \bibnamefont{Berg}},
  \bibinfo{journal}{Proc. Nat. Acad. Sci. U.S.A.}
  \textbf{\bibinfo{volume}{83}}, \bibinfo{pages}{8987} (\bibinfo{year}{1986}),
  ISSN \bibinfo{issn}{0027-8424}.

\bibitem[{\citenamefont{Kashikar et~al.}(2012)\citenamefont{Kashikar, Alvarez,
  Seifert, Gregor, J{\"a}ckle, Beyermann, Krause, and Kaupp}}]{Kashikar2012}
\bibinfo{author}{\bibfnamefont{N.~D.} \bibnamefont{Kashikar}},
  \bibinfo{author}{\bibfnamefont{L.}~\bibnamefont{Alvarez}},
  \bibinfo{author}{\bibfnamefont{R.}~\bibnamefont{Seifert}},
  \bibinfo{author}{\bibfnamefont{I.}~\bibnamefont{Gregor}},
  \bibinfo{author}{\bibfnamefont{O.}~\bibnamefont{J{\"a}ckle}},
  \bibinfo{author}{\bibfnamefont{M.}~\bibnamefont{Beyermann}},
  \bibinfo{author}{\bibfnamefont{E.}~\bibnamefont{Krause}}, \bibnamefont{and}
  \bibinfo{author}{\bibfnamefont{U.~B.} \bibnamefont{Kaupp}},
  \bibinfo{journal}{J. Cell Biol.} \textbf{\bibinfo{volume}{198}},
  \bibinfo{pages}{1075} (\bibinfo{year}{2012}).

\bibitem[{\citenamefont{Hathcock et~al.}(2016)\citenamefont{Hathcock, Sheehy,
  Weisenberger, Ilker, and Hinczewski}}]{Hathcock2016}
\bibinfo{author}{\bibfnamefont{D.}~\bibnamefont{Hathcock}},
  \bibinfo{author}{\bibfnamefont{J.}~\bibnamefont{Sheehy}},
  \bibinfo{author}{\bibfnamefont{C.}~\bibnamefont{Weisenberger}},
  \bibinfo{author}{\bibfnamefont{E.}~\bibnamefont{Ilker}}, \bibnamefont{and}
  \bibinfo{author}{\bibfnamefont{M.}~\bibnamefont{Hinczewski}},
  \bibinfo{journal}{IEEE Trans. Mol. Biol. Multi-Scale Commun.}
  \textbf{\bibinfo{volume}{2}}, \bibinfo{pages}{16} (\bibinfo{year}{2016}).

\bibitem[{\citenamefont{Celani and Vergassola}(2010)}]{Celani2010}
\bibinfo{author}{\bibfnamefont{A.}~\bibnamefont{Celani}} \bibnamefont{and}
  \bibinfo{author}{\bibfnamefont{M.}~\bibnamefont{Vergassola}},
  \bibinfo{journal}{Proc. Natl. Acad. Sci. U.S.A.}
  \textbf{\bibinfo{volume}{107}}, \bibinfo{pages}{1391} (\bibinfo{year}{2010}).

\bibitem[{\citenamefont{Aquino et~al.}(2014)\citenamefont{Aquino, Tweedy,
  Heinrich, and Endres}}]{Aquino2014}
\bibinfo{author}{\bibfnamefont{G.}~\bibnamefont{Aquino}},
  \bibinfo{author}{\bibfnamefont{L.}~\bibnamefont{Tweedy}},
  \bibinfo{author}{\bibfnamefont{D.}~\bibnamefont{Heinrich}}, \bibnamefont{and}
  \bibinfo{author}{\bibfnamefont{R.~G.} \bibnamefont{Endres}},
  \bibinfo{journal}{Sci. Rep.} \textbf{\bibinfo{volume}{4}}, \bibinfo{pages}{1}
  (\bibinfo{year}{2014}).

\bibitem[{\citenamefont{Hein et~al.}(2016)\citenamefont{Hein, Brumley, Carrara,
  Stocker, and Levin}}]{Hein2016}
\bibinfo{author}{\bibfnamefont{A.~M.} \bibnamefont{Hein}},
  \bibinfo{author}{\bibfnamefont{D.~R.} \bibnamefont{Brumley}},
  \bibinfo{author}{\bibfnamefont{F.}~\bibnamefont{Carrara}},
  \bibinfo{author}{\bibfnamefont{R.}~\bibnamefont{Stocker}}, \bibnamefont{and}
  \bibinfo{author}{\bibfnamefont{S.~A.} \bibnamefont{Levin}},
  \bibinfo{journal}{J. R. Soc. Interface} \textbf{\bibinfo{volume}{13}}
  (\bibinfo{year}{2016}).

\bibitem[{\citenamefont{Strong et~al.}(1998)\citenamefont{Strong, Freedman,
  Bialek, and Koberle}}]{Strong1998}
\bibinfo{author}{\bibfnamefont{S.~P.} \bibnamefont{Strong}},
  \bibinfo{author}{\bibfnamefont{B.}~\bibnamefont{Freedman}},
  \bibinfo{author}{\bibfnamefont{W.}~\bibnamefont{Bialek}}, \bibnamefont{and}
  \bibinfo{author}{\bibfnamefont{R.}~\bibnamefont{Koberle}},
  \bibinfo{journal}{Phys. Rev. E} \textbf{\bibinfo{volume}{57}},
  \bibinfo{pages}{4604} (\bibinfo{year}{1998}).

\bibitem[{\citenamefont{Andrews et~al.}(2006)\citenamefont{Andrews, Yi, and
  Iglesias}}]{Andrews2006}
\bibinfo{author}{\bibfnamefont{B.~W.} \bibnamefont{Andrews}},
  \bibinfo{author}{\bibfnamefont{T.~M.} \bibnamefont{Yi}}, \bibnamefont{and}
  \bibinfo{author}{\bibfnamefont{P.~A.} \bibnamefont{Iglesias}},
  \bibinfo{journal}{PLoS Comp. Biol.} \textbf{\bibinfo{volume}{2}},
  \bibinfo{pages}{1407} (\bibinfo{year}{2006}).

\bibitem[{\citenamefont{Kalman}(1960)}]{Kalman1960}
\bibinfo{author}{\bibfnamefont{R.~E.} \bibnamefont{Kalman}},
  \bibinfo{journal}{Trans. ASME -- J. Basic Engin.}
  \textbf{\bibinfo{volume}{82}}, \bibinfo{pages}{35} (\bibinfo{year}{1960}).

\bibitem[{\citenamefont{Melsa and Cohn}(1978)}]{Melsa1978}
\bibinfo{author}{\bibfnamefont{J.~L.} \bibnamefont{Melsa}} \bibnamefont{and}
  \bibinfo{author}{\bibfnamefont{D.~L.} \bibnamefont{Cohn}},
  \emph{\bibinfo{title}{Decision and estimation theory}}
  (\bibinfo{publisher}{McGraw-Hill, New York}, \bibinfo{year}{1978}).

\bibitem[{\citenamefont{Kobayashi}(2010)}]{Kobayashi2010}
\bibinfo{author}{\bibfnamefont{T.~J.} \bibnamefont{Kobayashi}},
  \bibinfo{journal}{Phys. Rev. Lett.} \textbf{\bibinfo{volume}{104}},
  \bibinfo{pages}{1} (\bibinfo{year}{2010}).

\bibitem[{\citenamefont{Endres and Wingreen}(2009)}]{Endres2009}
\bibinfo{author}{\bibfnamefont{R.~G.} \bibnamefont{Endres}} \bibnamefont{and}
  \bibinfo{author}{\bibfnamefont{N.~S.} \bibnamefont{Wingreen}},
  \bibinfo{journal}{Phys. Rev. Lett.} \textbf{\bibinfo{volume}{103}},
  \bibinfo{pages}{158101} (\bibinfo{year}{2009}).

\bibitem[{\citenamefont{Zechner et~al.}(2016)\citenamefont{Zechner, Seelig,
  Rullan, and Khammash}}]{Zechner2016}
\bibinfo{author}{\bibfnamefont{C.}~\bibnamefont{Zechner}},
  \bibinfo{author}{\bibfnamefont{G.}~\bibnamefont{Seelig}},
  \bibinfo{author}{\bibfnamefont{M.}~\bibnamefont{Rullan}}, \bibnamefont{and}
  \bibinfo{author}{\bibfnamefont{M.}~\bibnamefont{Khammash}},
  \bibinfo{journal}{Proc. Natl. Acad. Sci. U.S.A.}
  \textbf{\bibinfo{volume}{113}}, \bibinfo{pages}{4729} (\bibinfo{year}{2016}).

\bibitem[{\citenamefont{Scholz et~al.}(2017)\citenamefont{Scholz, Dinner,
  Levine, and Biron}}]{Scholz2017}
\bibinfo{author}{\bibfnamefont{M.}~\bibnamefont{Scholz}},
  \bibinfo{author}{\bibfnamefont{A.~R.} \bibnamefont{Dinner}},
  \bibinfo{author}{\bibfnamefont{E.}~\bibnamefont{Levine}}, \bibnamefont{and}
  \bibinfo{author}{\bibfnamefont{D.}~\bibnamefont{Biron}},
  \bibinfo{journal}{Proc. Natl. Acad. Sci. U.S.A.}
  \textbf{\bibinfo{volume}{114}}, \bibinfo{pages}{9261} (\bibinfo{year}{2017}).

\bibitem[{\citenamefont{Mora and Wingreen}(2010)}]{Mora2010}
\bibinfo{author}{\bibfnamefont{T.}~\bibnamefont{Mora}} \bibnamefont{and}
  \bibinfo{author}{\bibfnamefont{N.~S.} \bibnamefont{Wingreen}},
  \bibinfo{journal}{Phys. Rev. Lett.} \textbf{\bibinfo{volume}{104}},
  \bibinfo{pages}{248101} (\bibinfo{year}{2010}).

\bibitem[{\citenamefont{Libby et~al.}(2007)\citenamefont{Libby, Perkins, and
  Swain}}]{Libby2007}
\bibinfo{author}{\bibfnamefont{E.}~\bibnamefont{Libby}},
  \bibinfo{author}{\bibfnamefont{T.~J.} \bibnamefont{Perkins}},
  \bibnamefont{and} \bibinfo{author}{\bibfnamefont{P.~S.} \bibnamefont{Swain}},
  \bibinfo{journal}{Proc. Natl. Acad. Sci. U.S.A.}
  \textbf{\bibinfo{volume}{104}}, \bibinfo{pages}{7151} (\bibinfo{year}{2007}).

\bibitem[{\citenamefont{Siggia and Vergassola}(2013)}]{Siggia2013}
\bibinfo{author}{\bibfnamefont{E.~D.} \bibnamefont{Siggia}} \bibnamefont{and}
  \bibinfo{author}{\bibfnamefont{M.}~\bibnamefont{Vergassola}},
  \bibinfo{journal}{Proc. Natl. Acad. Sci. U.S.A.}
  \textbf{\bibinfo{volume}{110}}, \bibinfo{pages}{E3704}
  (\bibinfo{year}{2013}).

\bibitem[{\citenamefont{Mayer et~al.}(2019)\citenamefont{Mayer,
  Balasubramanian, Walczak, and Mora}}]{Mayer2019}
\bibinfo{author}{\bibfnamefont{A.}~\bibnamefont{Mayer}},
  \bibinfo{author}{\bibfnamefont{V.}~\bibnamefont{Balasubramanian}},
  \bibinfo{author}{\bibfnamefont{A.~M.} \bibnamefont{Walczak}},
  \bibnamefont{and} \bibinfo{author}{\bibfnamefont{T.}~\bibnamefont{Mora}},
  \bibinfo{journal}{Proc. Natl. Acad. Sci. U.S.A.}
  \textbf{\bibinfo{volume}{116}}, \bibinfo{pages}{8815} (\bibinfo{year}{2019}).

\bibitem[{\citenamefont{Hu et~al.}(2011)\citenamefont{Hu, Chen, Rappel, and
  Levine}}]{Hu2011}
\bibinfo{author}{\bibfnamefont{B.}~\bibnamefont{Hu}},
  \bibinfo{author}{\bibfnamefont{W.}~\bibnamefont{Chen}},
  \bibinfo{author}{\bibfnamefont{W.~J.} \bibnamefont{Rappel}},
  \bibnamefont{and} \bibinfo{author}{\bibfnamefont{H.}~\bibnamefont{Levine}},
  \bibinfo{journal}{Phys. Rev. E} \textbf{\bibinfo{volume}{83}},
  \bibinfo{pages}{021917} (\bibinfo{year}{2011}).

\bibitem[{\citenamefont{Vergassola et~al.}(2007)\citenamefont{Vergassola,
  Villermaux, and Shraiman}}]{Vergassola2007}
\bibinfo{author}{\bibfnamefont{M.}~\bibnamefont{Vergassola}},
  \bibinfo{author}{\bibfnamefont{E.}~\bibnamefont{Villermaux}},
  \bibnamefont{and} \bibinfo{author}{\bibfnamefont{B.}~\bibnamefont{Shraiman}},
  \bibinfo{journal}{Nature} \textbf{\bibinfo{volume}{445}},
  \bibinfo{pages}{406} (\bibinfo{year}{2007}).

\bibitem[{\citenamefont{Petrovi\'{c} and Markovi\'{c}}(2012)}]{Markovic2012}
\bibinfo{author}{\bibfnamefont{I.}~\bibnamefont{Petrovi\'{c}}}
  \bibnamefont{and}
  \bibinfo{author}{\bibfnamefont{I.}~\bibnamefont{Markovi\'{c}}}, in
  \emph{\bibinfo{booktitle}{2012 IEEE/RSJ International Conference on
  Intelligent Robots and Systems}} (\bibinfo{year}{2012}), pp.
  \bibinfo{pages}{707--712}.

\bibitem[{\citenamefont{{Kurz} et~al.}(2016)\citenamefont{{Kurz},
  {Gilitschenski}, and {Hanebeck}}}]{Kurz2016}
\bibinfo{author}{\bibfnamefont{G.}~\bibnamefont{{Kurz}}},
  \bibinfo{author}{\bibfnamefont{I.}~\bibnamefont{{Gilitschenski}}},
  \bibnamefont{and} \bibinfo{author}{\bibfnamefont{U.~D.}
  \bibnamefont{{Hanebeck}}}, \bibinfo{journal}{IEEE Aero. El. Sys. Mag.}
  \textbf{\bibinfo{volume}{31}}, \bibinfo{pages}{70} (\bibinfo{year}{2016}).

\bibitem[{\citenamefont{Kromer et~al.}(2018)\citenamefont{Kromer,
  M{\"{a}}rcker, Lange, Baier, and Friedrich}}]{Kromer2018}
\bibinfo{author}{\bibfnamefont{J.}~\bibnamefont{Kromer}},
  \bibinfo{author}{\bibfnamefont{S.}~\bibnamefont{M{\"{a}}rcker}},
  \bibinfo{author}{\bibfnamefont{S.}~\bibnamefont{Lange}},
  \bibinfo{author}{\bibfnamefont{C.}~\bibnamefont{Baier}}, \bibnamefont{and}
  \bibinfo{author}{\bibfnamefont{B.~M.} \bibnamefont{Friedrich}},
  \bibinfo{journal}{PLoS Comp. Biol.} \textbf{\bibinfo{volume}{14}},
  \bibinfo{pages}{1} (\bibinfo{year}{2018}).

\bibitem[{\citenamefont{Arkowitz}(1999)}]{Arkowitz1999}
\bibinfo{author}{\bibfnamefont{R.~A.} \bibnamefont{Arkowitz}},
  \bibinfo{journal}{Trends Cell Biol.} \textbf{\bibinfo{volume}{9}},
  \bibinfo{pages}{20} (\bibinfo{year}{1999}).

\bibitem[{\citenamefont{Manahan et~al.}(2004)\citenamefont{Manahan, Iglesias,
  Long, and Devreotes}}]{Manahan2004}
\bibinfo{author}{\bibfnamefont{C.~L.} \bibnamefont{Manahan}},
  \bibinfo{author}{\bibfnamefont{P.~A.} \bibnamefont{Iglesias}},
  \bibinfo{author}{\bibfnamefont{Y.}~\bibnamefont{Long}}, \bibnamefont{and}
  \bibinfo{author}{\bibfnamefont{P.~N.} \bibnamefont{Devreotes}},
  \bibinfo{journal}{Annu. Rev. Cell Dev. Biol.} \textbf{\bibinfo{volume}{20}},
  \bibinfo{pages}{223} (\bibinfo{year}{2004}).

\bibitem[{\citenamefont{Sarris and Sixt}(2015)}]{Sarris2015}
\bibinfo{author}{\bibfnamefont{M.}~\bibnamefont{Sarris}} \bibnamefont{and}
  \bibinfo{author}{\bibfnamefont{M.}~\bibnamefont{Sixt}},
  \bibinfo{journal}{Curr. Opin. Cell Biol.} \textbf{\bibinfo{volume}{36}},
  \bibinfo{pages}{93} (\bibinfo{year}{2015}).

\bibitem[{\citenamefont{Ma et~al.}(2014)\citenamefont{Ma, Klindt, Riedel-Kruse,
  J{\"u}licher, and Friedrich}}]{Ma2014}
\bibinfo{author}{\bibfnamefont{R.}~\bibnamefont{Ma}},
  \bibinfo{author}{\bibfnamefont{G.~S.} \bibnamefont{Klindt}},
  \bibinfo{author}{\bibfnamefont{I.~H.} \bibnamefont{Riedel-Kruse}},
  \bibinfo{author}{\bibfnamefont{F.}~\bibnamefont{J{\"u}licher}},
  \bibnamefont{and} \bibinfo{author}{\bibfnamefont{B.~M.}
  \bibnamefont{Friedrich}}, \bibinfo{journal}{Phys. Rev. Lett.}
  \textbf{\bibinfo{volume}{113}}, \bibinfo{pages}{048101}
  (\bibinfo{year}{2014}).

\bibitem[{\citenamefont{Friedrich}(2008)}]{Friedrich2008}
\bibinfo{author}{\bibfnamefont{B.~M.} \bibnamefont{Friedrich}},
  \bibinfo{journal}{Phys. Biol.} \textbf{\bibinfo{volume}{5}},
  \bibinfo{pages}{026007} (\bibinfo{year}{2008}).

\bibitem[{\citenamefont{Van~Haastert}(2010)}]{vanHaastert2010}
\bibinfo{author}{\bibfnamefont{P.~J.} \bibnamefont{Van~Haastert}},
  \bibinfo{journal}{PLoS Comp. Biol.} \textbf{\bibinfo{volume}{6}}
  (\bibinfo{year}{2010}).

\bibitem[{\citenamefont{Jerison and Quake}(2019)}]{Jerison2019}
\bibinfo{author}{\bibfnamefont{E.~R.} \bibnamefont{Jerison}} \bibnamefont{and}
  \bibinfo{author}{\bibfnamefont{S.~R.} \bibnamefont{Quake}},
  \bibinfo{journal}{bioRxiv}  (\bibinfo{year}{2019}),
  \urlprefix\url{https://www.biorxiv.org/content/early/2019/12/28/785964}.

\bibitem[{\citenamefont{Berg}(1993)}]{Berg1993}
\bibinfo{author}{\bibfnamefont{H.~C.} \bibnamefont{Berg}},
  \emph{\bibinfo{title}{Random Walks in Biology}}
  (\bibinfo{publisher}{Princeton Univ. Press}, \bibinfo{year}{1993}).

\bibitem[{\citenamefont{Str{\"u}nker et~al.}(2015)\citenamefont{Str{\"u}nker,
  Alvarez, and Kaupp}}]{Strunker2015}
\bibinfo{author}{\bibfnamefont{T.}~\bibnamefont{Str{\"u}nker}},
  \bibinfo{author}{\bibfnamefont{L.}~\bibnamefont{Alvarez}}, \bibnamefont{and}
  \bibinfo{author}{\bibfnamefont{U.}~\bibnamefont{Kaupp}},
  \bibinfo{journal}{Curr. Opinion Neurobiol.} \textbf{\bibinfo{volume}{34}},
  \bibinfo{pages}{110} (\bibinfo{year}{2015}).

\bibitem[{\citenamefont{Ueda and Shibata}(2007)}]{Ueda2007}
\bibinfo{author}{\bibfnamefont{M.}~\bibnamefont{Ueda}} \bibnamefont{and}
  \bibinfo{author}{\bibfnamefont{T.}~\bibnamefont{Shibata}},
  \bibinfo{journal}{Biophys. J.} \textbf{\bibinfo{volume}{93}},
  \bibinfo{pages}{11} (\bibinfo{year}{2007}).

\bibitem[{\citenamefont{Ma et~al.}(2004)\citenamefont{Ma, Janetopoulos, Yang,
  Devreotes, and Iglesias}}]{Ma2004}
\bibinfo{author}{\bibfnamefont{L.}~\bibnamefont{Ma}},
  \bibinfo{author}{\bibfnamefont{C.}~\bibnamefont{Janetopoulos}},
  \bibinfo{author}{\bibfnamefont{L.}~\bibnamefont{Yang}},
  \bibinfo{author}{\bibfnamefont{P.~N.} \bibnamefont{Devreotes}},
  \bibnamefont{and} \bibinfo{author}{\bibfnamefont{P.~A.}
  \bibnamefont{Iglesias}}, \bibinfo{journal}{Biophys. J.}
  \textbf{\bibinfo{volume}{87}}, \bibinfo{pages}{3764} (\bibinfo{year}{2004}).

\bibitem[{\citenamefont{Levine et~al.}(2006)\citenamefont{Levine, Kessler, and
  ~}}]{Levine2006}
\bibinfo{author}{\bibfnamefont{H.}~\bibnamefont{Levine}},
  \bibinfo{author}{\bibfnamefont{D.~A.} \bibnamefont{Kessler}},
  \bibnamefont{and} \bibinfo{author}{\bibfnamefont{W.-J.} \bibnamefont{~}},
  \bibinfo{journal}{Proc. Natl. Acad. Sci. U.S.A.}
  \textbf{\bibinfo{volume}{103}}, \bibinfo{pages}{9761} (\bibinfo{year}{2006}).

\bibitem[{\citenamefont{Skoge et~al.}(2014)\citenamefont{Skoge, Yue, Erickstad,
  Bae, Levine, Groisman, Loomis, and Rappel}}]{Skoge2014}
\bibinfo{author}{\bibfnamefont{M.}~\bibnamefont{Skoge}},
  \bibinfo{author}{\bibfnamefont{H.}~\bibnamefont{Yue}},
  \bibinfo{author}{\bibfnamefont{M.}~\bibnamefont{Erickstad}},
  \bibinfo{author}{\bibfnamefont{A.}~\bibnamefont{Bae}},
  \bibinfo{author}{\bibfnamefont{H.}~\bibnamefont{Levine}},
  \bibinfo{author}{\bibfnamefont{A.}~\bibnamefont{Groisman}},
  \bibinfo{author}{\bibfnamefont{W.~F.} \bibnamefont{Loomis}},
  \bibnamefont{and} \bibinfo{author}{\bibfnamefont{W.-J.}
  \bibnamefont{Rappel}}, \bibinfo{journal}{Proc. Natl. Acad. Sci. U. S. A}
  \textbf{\bibinfo{volume}{111}}, \bibinfo{pages}{14448}
  (\bibinfo{year}{2014}).

\bibitem[{\citenamefont{Andrews and Iglesias}(2007)}]{Andrews2007}
\bibinfo{author}{\bibfnamefont{B.~W.} \bibnamefont{Andrews}} \bibnamefont{and}
  \bibinfo{author}{\bibfnamefont{P.~A.} \bibnamefont{Iglesias}},
  \bibinfo{journal}{PLoS Comp. Biol.} \textbf{\bibinfo{volume}{3}},
  \bibinfo{pages}{1} (\bibinfo{year}{2007}).

\bibitem[{\citenamefont{Friedrich and J\"ulicher}(2009)}]{Friedrich2009}
\bibinfo{author}{\bibfnamefont{B.~M.} \bibnamefont{Friedrich}}
  \bibnamefont{and}
  \bibinfo{author}{\bibfnamefont{F.}~\bibnamefont{J\"ulicher}},
  \bibinfo{journal}{Phys. Rev. Lett.} \textbf{\bibinfo{volume}{103}},
  \bibinfo{pages}{068102} (\bibinfo{year}{2009}).

\bibitem[{\citenamefont{Mardia and Jupp}(2000)}]{Mardia2000}
\bibinfo{author}{\bibfnamefont{K.~V.} \bibnamefont{Mardia}} \bibnamefont{and}
  \bibinfo{author}{\bibfnamefont{P.~E.} \bibnamefont{Jupp}},
  \emph{\bibinfo{title}{Directional statistics}} (\bibinfo{publisher}{John
  Wiley \& Sons}, \bibinfo{year}{2000}).

\end{thebibliography}


\clearpage


\renewcommand{\theequation}{S\arabic{equation}}    
\setcounter{equation}{0}  
\renewcommand{\thefigure}{S\arabic{figure}}    
\setcounter{figure}{0}  
\renewcommand{\thetable}{S\arabic{table}}    
\setcounter{table}{0}  
\renewcommand{\thepage}{S\arabic{page}}  
\setcounter{page}{1}  
\renewcommand{\thesection}{\Alph{section}}
\renewcommand{\thesubsection}{\thesection.\arabic{subsection}}
\setcounter{section}{0}


\section{Details on analytical calculations}

\subsection{Expectation values of higher moments}
\label{app:expectation_values}

In deriving Eqs.~(\ref{eq:mean}) and (\ref{eq:covar})
for the mean $\mu_0$ and covariance matrix $\Sigma_0$ of a single measurement $\M$, 
valid for $N\ge 4$,
we used
\begin{equation}
\begin{split}
\label{eq:n0_n1}
\langle\wt{n}_0\rangle = \nu_0 \quad , \quad &
\langle\wt{n}_0^2\rangle = \langle\wt{n}_0\rangle^2 + \nu_0 \quad , \quad \\
\langle\wt{n}_1\rangle = \frac{\alpha_0\nu_0}{2} \, e^{i\psi_0} \quad , \quad &
\langle|\wt{n}_1|^2\rangle 
= | \langle\wt{n}_1 \rangle|^2 + \nu_0 \quad , \quad
\langle\wt{n}_0\wt{n}_1\rangle 
= \langle\wt{n}_0\rangle \langle\wt{n}_1\rangle + \langle\wt{n}_1\rangle \quad.
\end{split}
\end{equation}

Similarly, 
\begin{align}
\langle \wt{n}_0^2 |\wt{n}_1|^2 \rangle 
&=
\langle \wt{n}_0^2\rangle \langle |\wt{n}_1|^2 \rangle + 
\alpha_0^2\nu_0^3 + \alpha_0^2\nu_0^2 + 2\nu_0^2 + \nu_0 \quad.
\end{align}
For the special case $N=2$, we find different from Eq.~(\ref{eq:n0_n1})
\begin{align}
\langle \wt{n}_1\rangle=\alpha_0\nu_0 \cos\psi_0 \quad.  
\end{align}
For $N=3$, we find the same first moments as in Eq.~(\ref{eq:n0_n1}),
but different covariance matrix
\begin{align}
\Sigma_0^{(N=3)}
=
\Sigma_0^{(N\ge 4)} + 
\frac{\alpha_0\nu_0}{4}
\begin{pmatrix}
0 & 0 & 0 \\
0 & \phantom{-} \cos(\psi_0) &          -  \sin(\psi_0)\\
0 &          -  \sin(\psi_0) & \phantom{-} \cos(\psi_0)
\end{pmatrix}\quad ,
\end{align}
where $\Sigma_0^{(N\ge 4)}$ denotes the result for $N\ge 4$ from Eq.~(\ref{eq:covar}).
The mathematical reason is that 
the calculation of $\langle\wt{n}_1\rangle$ involves 
a sum of \textit{squared} roots of unity, $\sum_{j=1}^N \eta^{2j}$, 
which is nonzero for $N=2$, 
while the calculation of $\langle\wt{n}_1^2\rangle$ for $\Sigma_0$ involves
a sum of \textit{cubed} roots of unity, $\sum_{j=1}^N \eta^{3j}$, 
which is nonzero for $N=3$. 

\subsection{Expected precision of a single measurement $\langle\kappa_\tau\rangle$}
\label{app:kappa_tau}

In the double limit of high signal-to-noise ratio, $\SNR\gg 1$, 
and weak gradients, $\alpha_0\ll 1$, 
the factors in the definition of 
$\kappa_\tau = \wt{n}_0 |\wt{n}_1| A / \nu$
are (approximately) statistically independent; 
hence
$\langle \kappa_\tau \rangle 
\approx \langle \wt{n}_0 \rangle \langle |\wt{n}_1| \rangle \langle A \rangle / \langle \nu \rangle 
= \alpha_0 \langle |\wt{n}_1| \rangle$, 
since $\langle A\rangle\approx \alpha_0$.

To compute $\langle |\wt{n}_1| \rangle$, 
we use the law of cosines
\begin{equation}
c = |\wt{n}_1| = \sqrt{ a^2 + b^2 - 2 a b \cos\varphi } \text{ where }
a = |\langle\wt{n}_1\rangle| = \alpha_0 \nu_0 / 2 \text{ and }
b = |\wt{n}_1 - \langle\wt{n}_1\rangle| \quad .
\end{equation}
The isotropy of the covariance matrix in Eq.~(\ref{eq:covar})
implies that $\varphi$ is a uniformly distributed random angle with 
probability distribution $p(\varphi)=(2\pi)^{-1}$, 
while $b^2$ follows a $\chi^2$-distribution for 2 degrees of freedom 
(namely, $\wt{n}'_1{-}\langle \wt{n}'_1\rangle$ and $\wt{n}''_1{-}\langle\wt{n}''_1\rangle$),
hence
$p(b) = (2b/\nu_0) \exp( - b^2/\nu_0 )$.
Now,
$\langle c \rangle = \int_0^\infty \! db\, p(b) \oint_0^{2\pi} \! d\varphi\, p(\varphi)\,c(b,\varphi)$.
The first integration,
$\mathcal{I}(b) = \oint_0^{2\pi} \! d\varphi\, p(\varphi)\,c(b,\varphi)$,
results in an elliptic integral, 
which, however, can be well approximated by
\begin{equation}
\mathcal{I}(b) \approx 
\begin{cases}
\frac{\alpha_0 \nu_0}{2} \left[ 1 + \left(\frac{b}{\alpha_0\nu_0}\right)^2 \right] & b\le \alpha_0\nu_0/2 \\
b \left[ 1 + \left[\frac{\alpha_0\nu_0}{4b}\right)^2 \right] & b > \alpha_0\nu_0/2 
\end{cases} \quad .
\end{equation}
The second integration can now be easily done, yielding
\begin{equation}
\langle c \rangle = \frac{1}{\alpha_0} \left( \SNR + \frac{1}{2} + \ldots \right)\quad, 
\end{equation}
where the ellipses represents terms that decay exponentially fast for $\SNR\gg 1$.

\subsection{Prefactor in Eq.~(\ref{eq:one_measurement})}
\label{app:likelihood_prefactor}

The prefactor $\Lambda$ in Eq.~(\ref{eq:one_measurement}) is independent of $\psi$
and reads
\begin{align}
\Lambda = 
P(\M_{m+1}\,| & \,\M_{1:m})^{-1}\,
\frac{1}{ \sqrt{ (2\pi)^3 (2-\alpha^2) \nu^3 } } \cdot \notag \\
& \exp \left[
	-\frac{1}{2\nu}
		\left( \wt{n}_0^2\frac{2}{2-\alpha^2} - 2 \wt{n}_0\nu  + \nu^2 \right)
\right] 
\exp\left(
-\frac{ 4+\alpha^2 }{ 2 (2-\alpha^2 ) \nu} |\wt{n}_1|^2
\right) \quad.
\end{align}
In the limit of low signal-to-noise ratio, $\SNR \ll 1$, and $\alpha\approx\alpha_0$, this expression simplifies to
\begin{align}
\Lambda = P(\M_{m+1}\,|\,\M_{1:m})^{-1}\,
\frac{1}{ 4 (\nu\pi)^{3/2} } 
e^{-\frac{ (\nu-\wt{n}_0)^2 }{ 2 \nu } } \,
e^{-\frac{ |\wt{n}_1|^2}{ \nu } } + \O( \alpha_0^4 \nu_0^{1/2} ) \quad.
\end{align}

\subsection{Precision of sequential measurements}
\label{app:kappa_m}

We compute the second moment $\langle \kappa_m^2 \rangle$ 
of the measure of concentration of the marginal likelihood distribution $\L_m(\psi)$ 
of the estimated gradient angle $\psi$ 
after $m$ subsequent measurements, 
approximating said distribution by a von-Mises distribution.

Analogous to Eq.~(\ref{eq:wh_psi}), we have
\begin{equation}
\label{eq:wh_psi_m}
\kappa_m e^{i\wh{\psi}_m} 
= \kappa_0 e^{i\psi_0} + \sum_{j=1}^m \kappa_\tau^{(j)} e^{i\psi_\tau^{(j)}}, 
\end{equation}
where 
$\kappa_\tau^{(j)}$ is the measure of concentration of the $j^\mathrm{th}$ measurement $\M_j$,
and 
$\wt{n}_0^{(j)}$ and
$\wt{n}_1^{(j)} = |\wt{n}_1^{(j)}| e^{ i\psi_\tau^{(j)} }$ denote
the respective zeroth and first Fourier modes of molecule counts in $\M_j$.
In the absence of rotational diffusion, 
subsequent measurements are independent random variables, 
hence
\begin{equation}
\langle \wt{n}_{j,0}\wt{n}_1^{(j)} \wt{n}_{k,0}\wt{n}_1^{(k)}{}^\ast \rangle = 
\langle \wt{n}_{j,0}\wt{n}_1^{(j)} \rangle \langle \wt{n}_{k,0}\wt{n}_1^{(k)}\rangle^\ast \quad .
\end{equation}
Thus, we find analogous to Eq.~(\ref{eq:kappa_1})
\begin{subequations}
\begin{align}
\label{eq:kappa_m_calc_a}
\langle \kappa_m^2 \rangle 
&=
\kappa_0^2 +
\frac{\alpha_0^2}{\nu_0^2} 
\left[
    \sum_j \langle \wt{n}_0^{(j)}{}^2 |\wt{n}_1^{(j)}|^2 \rangle +
    \sum_{j\neq k} 
        \langle \wt{n}_0^{(j)} \wt{n}_1^{(j)} 
        \wt{n}_{k,0}\wt{n}_1^{(k)}{}^\ast \rangle
\right] +
\frac{\alpha_0}{\nu_0} 
\kappa_0 e^{i\psi_0} 
\sum_j \langle \wt{n}_0^{(j)} \wt{n}_1^{(j)}\rangle^\ast
+ \mathrm{c.c.} \\
\label{eq:kappa_m_calc_b}
&= 
\kappa_0^2 + 
\frac{\alpha_0^2}{\nu_0^2} 
\left[
\sum_j \langle \wt{n}_{j,0}^2 |\wt{n}_1^{(j)}|^2 \rangle + 
\sum_{j\neq k} 
    \langle \wt{n}_0^{(j)}\wt{n}_1^{(j)} \rangle
    \langle \wt{n}_0^{(k)}\wt{n}_1^{(k)} \rangle^\ast
\right] +
\frac{\alpha_0}{\nu_0} 
\kappa_0 e^{i\psi_0} \sum_j \langle \wt{n}_{j,0}\wt{n}_1^{(j)}\rangle^\ast +
\mathrm{c.c.} \\
\label{eq:kappa_m_calc_c}
&=
\kappa_0^2 + 
2(1+\kappa_0)\,m\SNR + 
m^2 \SNR^2 + 
\mathcal{O}(\alpha_0^2,\alpha_0^4\nu_0) \quad.
\end{align}
\end{subequations}
In the presence of rotational diffusion, $D>0$, 
Eq.~(\ref{eq:kappa_m_calc_a}) still holds, but Eq.~(\ref{eq:kappa_m_calc_b}) does not.
We use the approximation that $\wt{n}_0^{(j)}$ and $\wt{n}_1^{(j)}$ are approximately independent for each measurement,
hence
\begin{align}
\label{eq:kappa_m_calc_d}
\langle \kappa_m^2 \rangle 
&\approx
\kappa_0^2 +
\frac{\alpha_0^2}{\nu_0^2} 
\left[
    \sum_j \langle \wt{n}_0^{(j)}{}^2 |\wt{n}_1^{(j)}|^2 \rangle +
    \sum_{j\neq k} 
        \langle \wt{n}_0^{(j)} \wt{n}_0^{(k)} \rangle
		\langle \wt{n}_1^{(j)} \wt{n}_1^{(k)}{}^\ast \rangle
\right] +
\frac{\alpha_0}{\nu_0} 
\kappa_0 e^{i\psi_0} 
\sum_j \langle \wt{n}_0^{(j)} \wt{n}_1^{(j)}\rangle^\ast
+ \mathrm{c.c.} \quad.
\end{align}
We first compute the second sum of expectation values in Eq.~(\ref{eq:kappa_m_calc_d})
by evaluating a double geometric series
\begin{align}
\label{eq:n0n1r2_and_n0n1i2_average}
\sum_{j\neq k}
       \langle \wt{n}_1^{(j)} \wt{n}_1^{(k)}{}^\ast \rangle 
&=
\frac{\alpha_0^2\nu_0^2}{4} 2 \sum_{j < k} e^{-D |k-j|\tau} 
= 
\frac{\alpha_0^2\nu_0^2}{2} 
\left( 
	\frac{ m-1 }{ e^{D\tau} - 1 } -
	\frac{1-e^{-(m-1) D\tau}}{(e^{D\tau}-1)^2} \right) \quad. 
\end{align}
Similarly, we find for the first sum
\begin{equation}
\label{eq:n0n1r_average}
e^{ i\psi_0 } 
\sum_{j=1}^m \langle \wt{n}_1^{(j)} \rangle^\ast
= \alpha_0 \nu_0 \sum_{j=1}^m e^{-j D\tau} 
= \frac{\alpha_0 \nu_0}{2} \frac{1-e^{- m D\tau}}{e^{D\tau}-1} .
\end{equation}
By inserting 
Eqs.~(\ref{eq:n0n1r2_and_n0n1i2_average}) and (\ref{eq:n0n1r_average}), as well as Eq.~(\ref{eq:n0_n1}) into Eq.~(\ref{eq:kappa_m_calc_d}),
we obtain
\begin{equation}
\langle \kappa_m ^2\rangle
= 
\kappa_0^2 + 
2\left( 
	m + 
	\kappa_0 \frac{1-e^{- m D\tau}}{e^{D\tau}-1} 
\right) \SNR +
2\left( 
    m \frac{1}{e^{D\tau}-1} - 
	e^{D\tau} 
		\frac{1-e^{- m D\tau}}{\left(e^{D\tau}-1 \right)^2} + 
	\frac{m}{2} 
\right) \SNR^2 +
\O(\alpha_0^2,\alpha_0^4\nu_0) \quad, 
\end{equation}
from which Eq.~(\ref{eq:kappa_m_D2}) follows.

\subsection{Accuracy of estimated gradient direction for agents not aware of motility noise}
\label{app:sigma2_agent_unaware}

We derive the asymptotic scaling of the variance parameter $\sigma^2_m$ 
characterizing the distribution of estimation errors 
$\delta_m = \wh{\psi}_m - \psi_m$
in section \ref{sec:presence_of_D}.
We start with the ansatz $\sigma^2_m=(\gamma m)^{1/2}+\O(1)$ for $m$ large.
Inserting
$\wh{\sigma}^2_m \approx \langle\kappa_m^2\rangle^{-1/2}$ from Eq.~(\ref{eq:kappa_m3}) and 
$\sigma^2_\tau \approx \langle\kappa_\tau\rangle^{-1}\approx \SNR^{-1}$ from Eq.~(\ref{eq:kappa_tau_high_SNR})
into Eq.~(\ref{eq:iteration_maxilik}), we obtain a self-consistency condition
\begin{align}
\sqrt{\gamma\,m}&=\frac{1}{\sqrt{\frac{D\,\tau}{2(m-1)}}+1}
\left(\sqrt{\gamma\,(m-1)}+2D\,\tau\right)
+\frac{\sqrt{\frac{D\,\tau}{2(m-1)}}}{\sqrt{\frac{D\,\tau}{2(m-1)}}+1}\SNR^{-1} \quad.
\end{align}
We expand left-hand and right-hand side of this equation into powers of $m^{-1/2}$, 
and match both the leading-order term $\O(m^{1/2})$ and the first-order correction $\O(1)$:
this yields 
$\gamma=2D\,\tau$
and validates the Ansatz.

\section{Basic properties of circular distributions}
\label{app:circular_distributions}


A probability distribution $p(\psi)$ of angles
should be $2\pi$-periodic, i.e., $p(\psi)=p(\psi+2\pi)$, and 
normalized to one on the unit circle, 
i.e., $\oint_0^{2\pi}\!d\psi\, p(\psi)=1$.
The \textit{circular variance} of such a circular distribution is defined as
\begin{equation}
\CV[p(\psi)] = 1 - \left| \oint_0^{2\pi}\!d\psi\, e^{i\psi} p(\psi) \right| \quad .
\end{equation}
An important circular distribution is the \textit{wrapped normal distribution}
\begin{equation}
\WN(\psi;\mu,\sigma^2) = \sum_{l=-\infty}^\infty (2\pi\sigma^2)^{-1/2}\,\exp\left( - \frac{(\psi-\mu+2\pi l)^2}{2\sigma^2} \right)
\end{equation}
with variance parameter $\sigma^2$, 
whose circular variance reads $\CV=1-e^{-\sigma^2/2}$.
In the limit $\sigma^2\ll 1$, $\CV\approx \sigma^2/2$.
The wrapped normal distribution is closely approximated by the von-Mises distribution, 
which is commonly used in directional statistics due to its mathematical tractability 
\cite{Mardia2000} 
\begin{equation}
\VM(\psi;\mu,\kappa) = (2\pi I_0(\kappa))^{-1}\, \exp \left[ \kappa \cos(\psi-\mu) \right] \quad ,
\end{equation}
where
$\kappa$ is the so-called \textit{measure of concentration} or precision, and
$I_n(\kappa)$ is the modified Bessel function of order $n$.
The circular variance reads 
$\CV = 1-I_1(\kappa)/I_0(\kappa)=1/(2\kappa)+1/(8\kappa^2)+\mathcal{O}(\kappa^{-3})$.

The normalized product of two von-Mises distributions, 
say $\VM(\psi;\mu_1,\kappa_1)$ and $\VM(\psi;\mu_2,\kappa_2)$,
is again a von-Mises distribution $\VM(\psi;\mu,\kappa)$.
Such normalized product appears, e.g., in Bayes formula, Eq.~(\ref{eq:Bayes_rule}).
Specifically, the mean $\mu$ and measure of concentration $\kappa$ of the normalized product satisfy
\begin{equation}
\label{eq:kappa_sum}
\kappa e^{i\mu} = \kappa_1 e^{i\mu_1} + \kappa_2 e^{i\mu_2} \quad.
\end{equation}
If mean values are close, $|\mu_1-\mu_2|\ll 1$, 
we have the approximate sum rule
$\kappa \approx \kappa_1 + \kappa_2$.

In contrast, the circular convolution of two von-Mises distributions 
$\VM(\psi;\mu_1,\kappa_1)$ and $\VM(\psi;\mu_2,\kappa_2)$
is only approximately a von-Mises distribution \cite{Mardia2000}.
To find such approximation,
one can first map the two von-Mises distributions onto wrapped normal distributions of same respective mean and circular variance,
and compute the convolution of these wrapped normal distributions \cite{Kurz2016}.
The convolution of two wrapped normal distributions,
say with variance parameters $\sigma_1^2$ and $\sigma_2^2$, 
is again a wrapped normal distribution with new variance parameter $\sigma^2=\sigma_1^2+\sigma_2^2$.
Finally, this new wrapped normal distribution $\WN(\psi;\mu_1+\mu_2,\sigma^2)$ is mapped back onto a von-Mises distribution $\VM(\psi;\mu,\kappa)$.
Thus,
$\VM(\psi;\mu,\kappa) \approx \VM(\psi;\mu_1,\kappa_1) * \VM(\psi;\mu_2,\kappa_2)$
with
$\mu=\mu_1+\mu_2$
and
$
I_1(\kappa) / I_0(\kappa) = \exp(-\sigma^2/2) = \exp(-\sigma_1^2) \exp(-\sigma_2^2) 
= [ I_1(\kappa_1) / I_0(\kappa_1) ] \cdot [I_1(\kappa_2) / I_0(\kappa_2) ]
$.
In the limit $\kappa_1,\kappa_2\gg 1$, we have
$\kappa^{-1} = \kappa_1^{-1} + \kappa_2^{-1}$. 

For ease of reference, we highlight the sum rules for the measure of concentration
of either a normalized product or convolution of two von-Mises distributions
\begin{align}
\label{eq:VM_conv}
\text{convolution: }   \quad \frac{1}{\kappa} &\approx \frac{1}{\kappa_1} + \frac{1}{\kappa_2} \quad, \\
\label{eq:VM_mult}
\text{multiplication: }\quad \kappa           &\approx \kappa_1 + \kappa_2 \quad. 
\end{align}
While Eq.~(\ref{eq:VM_conv}) is valid for 
$\kappa_1,\kappa_2\gg 1$, 
Eq.~(\ref{eq:VM_mult}) is valid for $|\mu_1-\mu_2|\ll 1$.
Note that Eq.~(\ref{eq:VM_mult}) is not used until section \ref{sec:agents_aware}; 
before we always use the exact expression Eq.~(\ref{eq:kappa_sum}).

\section{Optimal averaging time for estimation of a scalar quantity}
\label{app:filter}

Previous work addressed the optimal time span for temporal averaging 
for chemotaxis by temporal comparison \cite{Segall1986}.
In our notation, this amounts to estimating the scalar quantity
$
s_\mathrm{true}(t) = \frac{d}{dt} c(\R(t)) =
v_0 \, \h_1\cdot\mathbf{\nabla}c
$
from a noisy input signal $s(t)$,
where the agent moves with velocity
$\dot{\R} = v_0 \h_1$.

As a minimal pedagogical model, 
we approximate $\true{s}(t)$ as an Ornstein-Uhlenbeck process with 
$\langle \true{s}(t) \rangle=0$ and
$\langle \true{s}(t) \true{s}(t-\Delta t) \rangle = (\alpha_0 c_0 v_0)^2/2 \, \exp(-D |\Delta t|)$.
We assume a measurement process with additive sensing noise, $s(t)=s_0(t)+\xi(t)$, 
where $\xi(t)$ denotes Gaussian white noise with
$\langle \xi(t) \rangle = 0$ and 
$\langle \xi(t)\xi(t') \rangle = (\sigma^2/\tau)\, \delta(t-t')$.
The agent shall perform temporal averaging using a linear filter $\chi(\Delta t)$
\begin{align}
\wh{s}(t)=\int_0^\infty \! d\Delta t\, s(t-\Delta t)\chi(\Delta t) \quad.
\end{align}
We restrict ourselves to filters of exponential form,
$\chi(\Delta) = A \exp(-\Delta t / t_\chi)$, 
and ask for the optimal averaging time span $t_\chi$. 
From the condition for a faithful estimtor, 
$\langle \wh{s}(t)\rangle = \true{s}(t)$, 
we obtain the prefactor $A$ as $A = D+1/t_\chi$
(where this and subsequent expectation values are conditioned on $\true{s}(t)$).
We minimize the variance of estimation errors 
$\delta(t) = \wh{s}(t) - \true{s}(t)$, 
which is equivalent to minimizing 
$\langle \wh{s}^2 \rangle$.
Using the autocorrelation function of $\true{s}(t)$ above, we find
\begin{align}
\langle \wh{s}^2 \rangle
= A^2
\int_0^\infty \int_0^\infty \! d\Delta t_1 d\Delta t_2 \,
\true{s^2}(t) e^{-D|\Delta t_1-\Delta t_2|} \, e^{-(\Delta t_1+\Delta t_2)/ t_\chi } \quad .
\end{align}
We compute this integral using the change of variables 
$z_1=\Delta t_1+\Delta t_2$ and $z_2=\Delta t_1-\Delta t_2$, 
and find
\begin{align}
\langle \wh{s}^2 \rangle = 
\true{s^2}(t) \,
\left(D+\frac{1}{t_\chi}\right)^2\,\frac{1}{D}\,\,t_\chi
\sim D t_\chi + 2 + \frac{1}{D t_\chi} 
\quad . 
\end{align}
This expectation value becomes minimal 
exactly for $t_\chi=D^{-1}$, 
irrespective of the value of $\true{s}(t)$.
Thus, the optimal averaging time equals the rotational diffusion time for estimating this scalar quantity. 
Similar results were found for detailed models of bacterial chemotaxis \cite{Segall1986}.
Note that in this case, the time-derivative $d/dt$ is incorporated into the filter function itself, representing a smoothed time derivative.

\section{Intermediate measurements give (almost) no advantage}
\label{app:no_impatience}

As in the main text, 
we consider an agent that monitors a static environment with constant state $\S(t)=S_0$,
where the agent takes subsequent measurements $\M_j$ during time intervals $\T_j=((j-1)\tau,j\tau)$.
We ask if knowing the results $\M_j$ of the intermediate measurements
confers any advantage compared to a single measurement of same total duration $m\tau$, 
corresponding to the sum $\M=\sum_{j=1}^m \M_j$, 
if rotational diffusion is absent, $D=0$.

The answer is `no' for a Poisson point process, 
e.g.\ if the agent estimates an absolute concentration $\S=c$ 
by counting the number of independent molecular binding events $\M_j$ in a time-interval $\T_j$.
We sketch the known proof for $m=2$.
We want to show
\begin{equation}
\label{eq_no_impatience}
\L(\S\,|\,\M_1,\M_2;\tau) = \L(\S\,|\,\M_1+\M_2;2\tau),
\end{equation}
where the left-hand and right-hand side denote the likelihood of state $\S$
estimated from two intermediate measurements $\M_1$ and $\M_2$, each of duration $\tau$,
or a single measurement of duration $2\tau$ given by $\M_+ = \M_1 + \M_2$, respectively. 
We assume that measurements are Poisson distributed, i.e.,
$P(\M|\S;\tau) = e^{-\mu} \mu^\M / \M!$,
where the mean number of binding events $\mu = \lambda c \tau$ 
within a time-interval $\tau$ is proportional to concentration $c$.
A straight-forward application of the Binomial formula yields
$ P(\M_+\,|\,\S;2\tau) = P(\M_1\,|\,\S;\tau) P(\M_2\,|\,\S;\tau) $
for the measurement probabilities.
Bayes' theorem implies Eq.~(\ref{eq_no_impatience}) for arbitrary prior $\L_0(\S)$.
The case of $k\ge 2$ measurements follows by induction.

In the limit of large molecular counts $\mu\gg 1$, 
the Poisson distribution is well approximated by a normal distribution. 
One may thus assume that a statement similar to Eq.~\ref{eq_no_impatience} also holds true 
for a Gaussian measurement model with
$P(\M_j\,|\,\mu;\tau) = \N(\M_j;\mu,\sigma^2)$, $j=1,2$, and
$P(\M_+\,|\,\mu;\tau) = \N(\M_+;2\mu,2\sigma^2)$, 
where $\mu$ and $\sigma^2$ are unknown parameters.
(As above, $\N(x,\mu,\sigma^2)$ denotes the normal distribution 
with argument $x$, mean $\mu$, variance $\sigma^2$.)
Intriguingly, the answer now depends on whether $\mu$ and $\sigma^2$ are independent or not, 
according to the Bayesian prior $\L_0(\S)=\L_0(\mu,\sigma^2)$.
If $\mu$ and $\sigma^2$ are independent, 
Eq.~(\ref{eq_no_impatience}) holds also for Gaussian measurements.
If, however, $\mu$ and $\sigma^2$ are independent, 
say $\sigma=\sigma(\mu)$ is a function of $\mu$,
an explicit calculation yields
\begin{equation}
\label{eq_no_impatience_normal}
    \L(\wh{\mu}\,|\,\M_1,\M_2;\tau) = 
    \mathcal{N}(\M_{-};0,2\sigma(\wh{\mu})^2)\,
    \L(\wh{\mu}\,|\,\M_1+\M_2;2\tau)\, .
\end{equation}
In short, knowledge of $\M_{-}=\M_2-\M_1$ improves the estimate for $\sigma^2$, 
which, in turn, improves the estimate $\wh{\mu}$ of $\mu$.
In the limit $d\sigma/d\mu \ll 1$, 
the pre-factor on the right-hand side of Eq.~(\ref{eq_no_impatience_normal})
will be approximately constant, 
displaying only relative changes on the order of 
$[\M_{-}/\sigma(\mu)]^2\,d\sigma/d\wh{\mu}_{|\wh{\mu}=\mu}$,
which, with probability close to $1$, is small.
Specifically, this will hold true 
for a diffusion approximation of a Poissonian measurement model, 
where $\hat{\mu}=\sigma^{2}(\hat{\mu})$, 
provided $\hat{\mu}$ is large, 
corresponding to the limit where Poissonian and Gaussian measurement models converge to each other.

The above result generalizes in a straight-forward manner to the case 
vectorial measurements that are distributed according to a multi-variate normal distribution
(by diagonalizing the co-variance matrix).
As a corollary, we thus obtain an analogous argument for the estimate of the direction of a concentration gradient, 
see also Eq.~(\ref{eq:kappa_m}).

In conclusion, 
taking intermediate measurements confers a minute advantage for gradient estimation, 
which vanishes in the limit of large molecular counts.



\end{document}